\documentclass[12pt,draftcls,onecolumn]{IEEEtran}

\usepackage{enumerate}
\usepackage{footnote} 
\usepackage{url}
\usepackage{cite}
\usepackage{epsfig}
\usepackage{subfigure}
\usepackage{amsmath,amssymb,amsfonts,mathrsfs,amscd}
\usepackage{flushend}

\newtheorem{proposition}{Proposition}
\newtheorem{remark}{Remark}
\newtheorem{definition}{Definition}
\newtheorem{theorem}{Theorem}

\newtheorem{corollary}{Corollary}

\newtheorem{lemma}{Lemma}

\newcommand{\bb}[1]{\mathbb{#1}}
\newcommand{\nnb}{\nonumber}

\newcommand{\Fig}[1]{Fig.~\!\ref{#1}}

\newcommand{\Lemma}[1]{lemma~\!\ref{#1}}
\newcommand{\Thm}[1]{theorem~\!\ref{#1}}
\newcommand{\Prop}[1]{proposition~\!\ref{#1}}
\newcommand{\App}[1]{Appendix~\!\ref{#1}}
\newcommand{\Eq}[1]{(\ref{#1})}

\newcommand{\ie}{\emph{i.e.}}
\newcommand{\eg}{\emph{e.g.}}
\newcommand{\vs}{\emph{vs. }}

\newcommand{\etal}{\emph{et al.}\,}

\newcommand{\for}{\textrm{for} }
\newcommand{\D}{\displaystyle}

\newcommand{\yseqncases}[5]{
  \begin{equation}
    \setlength{\nulldelimiterspace}{0pt}
    #1=\left\{
      \begin{IEEEeqnarraybox}[\relax][c]{l's}
        #2, &for #3\\
        #4, &for #5%
      \end{IEEEeqnarraybox}\right.
  \end{equation}
}

\newcommand{\yseqncasesasymptnnb}[5]{
  \begin{equation*}
    \setlength{\nulldelimiterspace}{0pt}
    #1\asympteq\left\{
      \begin{IEEEeqnarraybox}[\relax][c]{l's}
        #2, &for #3\\
        #4, &for #5%
      \end{IEEEeqnarraybox}\right.
  \end{equation*}
}

\renewcommand{\matrix}[1]{\begin{bmatrix}#1\end{bmatrix}}

\newcommand{\sss}{\scriptscriptstyle}
\newcommand{\mbs}[1]{\boldsymbol{#1}}
\renewcommand{\mbs}[1]{\mathbf{#1}}
\renewcommand{\mbs}[1]{\pmb{#1}}
\renewcommand{\th}{\textrm{th}}
\newcommand{\vect}[1]{{\lowercase{\mbs{#1}}}}
\newcommand{\mat}[1]{{\uppercase{\mbs{#1}}}}
\newcommand{\wt}{\widetilde}
\newcommand{\wh}{\widehat}

\newcommand\transcsymbol{\scriptscriptstyle \dag \!}
\newcommand\transsymbol{\scriptscriptstyle \mathsf{T} \!}

\newcommand\Abs[1]{\left|#1\right|}

\newcommand\Abssqr[1]{\left|#1\right|^2}

\newcommand{\inv}[1]{{#1}^{\scriptscriptstyle -1 \!}}
\newcommand{\transc}[1]{{#1}^{\transcsymbol}}
\newcommand{\trans}[1]{{#1}^{\transsymbol}}

\newcommand{\ssT}{{\transsymbol}}
\newcommand\diag{\mathrm{diag}}
\newcommand\Tr{\mathrm{Tr}}
\newcommand\Norm[1]{\left\|{#1}\right\|}
\newcommand\Frob[1]{\Norm{#1}^2_{\textrm{F}}}

\newcommand\defeq{\triangleq}

\newcommand\Prob[1]{\textrm{Prob}\left\{#1\right\}}
\newcommand\prob{\textrm{Prob}}

\newcommand\OK{\Ocal_{\KK}}

\newcommand{\Id}{\mathbf{I}}
\newcommand{\CN}[1][\Id]{\Ccal\Ncal\!\left(0,#1\right)}

\newcommand{\SNR}{{\mathsf{SNR}} }
\newcommand{\FER}{\textrm{FER}}

\newcommand\iid{i.i.d.\ }

\newcommand\Resp{\emph{resp.}\ }

\newcommand\Pe{P_{\textrm{e}}}
\newcommand\Pout{P_{\textrm{out}}}

\newcommand\One{\mathbf{1}}
\newcommand\mZero{\mathbf{0}}
\renewcommand\d{\mathrm{d}}

\newcommand\CC{\bb{C}}
\newcommand\EE{\bb{E}}

\newcommand\KK{\bb{K}}

\newcommand\RR{\bb{R}}

\newcommand\Acal{\mathcal{A}}
\newcommand\Bcal{\mathcal{B}}
\newcommand\Ccal{\mathcal{C}}

\newcommand\Ecal{\mathcal{E}}
\newcommand\Fcal{\mathcal{F}}

\newcommand\Ical{\mathcal{I}}

\newcommand\Ncal{\mathcal{N}}
\newcommand\Ocal{\mathcal{O}}

\newcommand\Scal{\mathcal{S}}

\newcommand\Xcal{\mathcal{X}}

\newcommand{\mA}{\mat{A}}
\newcommand{\mB}{\mat{B}}
\newcommand{\mC}{\mat{C}}
\newcommand{\mD}{\mat{D}}

\newcommand{\mF}{\mat{F}}
\newcommand{\mG}{\mat{G}}
\newcommand{\mH}{\mat{H}}
\newcommand{\mI}{\mat{I}}

\newcommand{\mL}{\mat{L}}
\newcommand{\mM}{\mat{M}}

\newcommand{\mP}{\mat{P}}

\newcommand{\mR}{\mat{R}}

\newcommand{\mU}{\mat{U}}
\newcommand{\mV}{\mat{V}}
\newcommand{\mW}{\mat{W}}
\newcommand{\mX}{\mat{X}}
\newcommand{\mY}{\mat{Y}}
\newcommand{\mZ}{\mat{Z}}

\newcommand{\mBtrc}[1][]{\mat{B}^{\transcsymbol}_{#1}}

\newcommand{\mGtrc}[1][]{\mat{G}^{\transcsymbol}_{#1}}
\newcommand{\mHtrc}[1][]{\mat{H}^{\transcsymbol}_{#1}}

\newcommand{\mPtrc}[1][]{\mat{P}^{\transcsymbol}_{#1}}

\newcommand{\ma}{\vect{a}}

\newcommand{\muu}{\vect{u}}
\newcommand{\mv}{\vect{v}}
\newcommand{\mw}{\vect{w}}
\newcommand{\mx}{\vect{x}}
\newcommand{\my}{\vect{y}}
\newcommand{\mz}{\vect{z}}

\newcommand{\mXi}{\mbs{\Xi}}

\newcommand{\mGamma}{\mbs{\Gamma}}
\newcommand{\mOmega}{\mbs{\Omega}}

\newcommand{\mSigma}{\mbs{\Sigma}}

\newcommand{\mLambda}{\mbs{\Lambda}}
\newcommand{\mlambda}{\boldsymbol{\lambda}}
\newcommand{\malpha}{\boldsymbol{\alpha}}
\newcommand{\mtheta}{\boldsymbol{\theta}}
\newcommand{\mmu}{\boldsymbol{\mu}}
\newcommand{\mbeta}{\boldsymbol{\beta}}

\newcommand{\NC}{\newcommand}
\newcommand{\RNC}{\renewcommand}

\NC{\He}{\mbs{\wt{H}}}
\NC{\mSigmasqrt}{\mGamma}
\NC{\mSigmasqrttrc}{\transc{\mSigmasqrt}}
\NC{\mSigmainv}{\mSigma^{-{1}}}
\NC{\AF}{\!\sss\Acal\Fcal}
\RNC{\AF}{\sss\textrm{AF}}

\newcommand{\ns}{n_{\textrm{s}}}
\newcommand{\nr}{n_{\textrm{r}}}
\newcommand{\nd}{n_{\textrm{d}}}
\newcommand{\nR}{n_{\textrm{R}}}
\newcommand{\nT}{n_{\textrm{T}}}

\NC{\Ixy}{\Ical(\mx;\sqrt{\SNR}\He\mx+\mz)}

\newcommand{\nntcode}{rate-$n$ NVD code}
\newcommand{\NAF}{\sss\textrm{NAF}} 
\newcommand{\CNAF}{C_{\NAF,N}}
\newcommand{\dX}{d_{\Xcal}}

\newcommand{\asympteq}{\doteq} %

\newcommand{\asymptleq}{\ \dot{\leq}\,}
\newcommand{\asymptgeq}{\ \dot{\geq}\,}
\newcommand{\asymptsubseteq}{\ \dot{\subseteq}\,}
\newcommand{\asymptsupseteq}{\ \dot{\supseteq}\,}

\newcommand{\Oe}{\Ocal}
\newcommand{\Oc}{\bar{\Ocal}}

\NC{\OneO}[2]{\One_{\Oe_{#1#2}}}
\NC{\Posr}{P_{\Oe_{sr}}}
\NC{\Poo}{P_{\Oe_{12},\Oe_{21}}}
\NC{\Pnoo}{P_{\bar{\Oe}_{12},\Oe_{21}}}
\NC{\Pono}{P_{\Oe_{12},\bar{\Oe}_{21}}}
\NC{\Pnono}{P_{\bar{\Oe}_{12},\bar{\Oe}_{21}}}
\NC{\out}{\textrm{out}}
\NC{\dHout}{d_{\mH}^{\out}}
\NC{\dout}{d_{\textrm{out}}}
\NC{\doutcc}{d^{\textrm{CC}}_{\textrm{out}}}
\NC{\doutnaf}{d^{\textrm{NAF}}_{\textrm{out}}}
\NC{\doutndf}{d^{\textrm{NDF}}_{\textrm{out}}}
\NC{\doutddf}{d^{\textrm{DDF}}_{\textrm{out}}}
\NC{\doutaf}{d^{\textrm{AF}}_{\textrm{out}}}
\NC{\doutdf}{d^{\textrm{DF}}_{\textrm{out}}}
\NC{\lmax}{\lambda_{\max}}
\NC{\Po}{P_{\Oe}}
\NC{\PE}{P_{\Ecal}}
\NC{\PEH}{P_{\Ecal_{\mH}}}
\NC{\PHH}{p_{\mH}(\mH)}
\NC{\PAA}{p_{\malpha}(\malpha)}
\NC{\dEmin}{d^2_{\min}}
\NC{\dEH}{d_{\Ecal|\mH}}

\NC{\Peo}{P_{\textrm{e},\Oe}}
\NC{\Peoc}{P_{\textrm{e},\Oc}}
\NC{\Pecondo}{P_{\textrm{e}|\Oe}}
\NC{\Pecondoc}{P_{\textrm{e}|\Oc}}
\NC{\Pepoc}{P_{\textrm{pe},\Oc}}
\NC{\Pepcondoc}{P_{\textrm{pe}|\Oc}}

\NC{\Tx}[1]{\textrm{Tx}(#1)}
\NC{\Rx}[1]{\textrm{Rx}(#1)}

\NC{\mHt}{\mbs{\wt H}}
\NC{\mHh}{\mbs{\wh H}}
\NC{\myt}{\mbs{\wt y}}
\NC{\mxt}{\mbs{\wt x}}
\NC{\mvt}{\mbs{\wt v}}
\NC{\mzetat}{\mbs{\wt \zeta}}
\NC{\mnt}{\mbs{\wt n}}
\NC{\mwt}{\mbs{\wt w}}
\NC{\mzt}{\mbs{\wt z}}

\NC{\range}[3]{\left\{#1\right\}^{#3}_{#2}}

\title{Optimal Space-Time Codes for the MIMO Amplify-and-Forward
  Cooperative Channel \thanks{Manuscript submitted to the IEEE
    Transactions on Information Theory.  S.~Yang and J.-C.~Belfiore
    are with the Department of Communications and Electronics,
    \'{E}cole Nationale Sup\'{e}rieure des T\'{e}l\'{e}communications,
    75013 Paris, France~(e-mail: syang@enst.fr; belfiore@enst.fr).}}

\author{Sheng Yang and Jean-Claude Belfiore}

\begin{document}
\maketitle

\begin{abstract}
  In this work, we extend the non-orthogonal amplify-and-forward~(NAF)
  cooperative diversity scheme to the MIMO channel. A family of
  space-time block codes for a half-duplex MIMO NAF fading cooperative
  channel with $N$ relays is constructed. The code construction is
  based on the non-vanishing determinant~(NVD) criterion and is shown
  to achieve the optimal diversity-multiplexing tradeoff~(DMT) of the
  channel. We provide a general explicit algebraic construction,
  followed by some examples. In particular, in the single-relay case,
  it is proved that the Golden code and the $4\times4$ Perfect code
  are optimal for the single-antenna and two-antenna case,
  respectively. Simulation results reveal that a significant gain~(up
  to $10$\,dB) can be obtained with the proposed codes, especially in
  the single-antenna case.
\end{abstract}


\section*{Notations} 
In this paper, we use boldface lower case letters $\mbs{v}$ to denote
vectors, boldface capital letters $\mbs{M}$ to denote matrices.
$\Ccal\Ncal$ represents the complex Gaussian random variable.
$\EE[\cdot]$ stands for the expectation operation and
$\trans{[\cdot]},\transc{[\cdot]}$ denote the matrix transposition and
conjugated transposition operations. $\Norm{\cdot}$ is the Euclidean
vector norm and $\Norm{\cdot}_{\text{F}}$ is the Frobenius matrix
norm.  $\Abs{\Scal}$ is the cardinality of the set $\Scal$.  $(x)^+$
means $\max(0,x)$. $\bb{R}$, $\bb{C}$, $\bb{Q}$ and $\bb{Z}$ stand for
the real field, complex field, rational field and the integer ring
respectively. For any quantity $q$,
  \begin{equation*}
    q \asympteq \SNR^{\alpha}\quad \textrm{means}\quad \lim_{\SNR\to\infty}\frac{\log q}{\log \SNR} = \alpha
  \end{equation*}
  and similarly for $\asymptleq$ and $\asymptgeq$.

\section{Introduction}
\label{sec:intro}
On a wireless channel, diversity techniques are used to combat channel
fadings. Recently, there has been a growing interest in the so called
\emph{cooperative diversity} techniques, where multiple terminals in a
network cooperate to form a virtual antenna array in order to exploit
spatial diversity in a distributed fashion. In this manner, spatial
diversity gain can be obtained even when a local antenna array is not
available. Since the work of \cite{Sendonaris1, Sendonaris2}, several
cooperative transmission protocols have been proposed~\cite{LTW1,
  LTW2, Nabar, ElGamal_coop, Hunter1, Hunter2, Tarokh_coop}. These
protocols can be categorized into two principal classes~: the
amplify-and-forward~(AF) scheme and the decode-and-forward~(DF)
scheme. In practice, the AF scheme is more attractive for its low
complexity since the cooperative terminals~(relays) simply forward the
signal and do not decode it. Actually, for most {\it ad hoc} wireless
networks, it is not realistic for other terminals to decode the signal
from a certain user, because the codebook is seldom available and the
decoding complexity is unacceptable in most cases.

The non-orthogonal amplify-and-forward~(NAF) scheme was proposed by
Nabar \etal~\cite{Nabar} for the single-relay channel and was then
generalized to the multiple-relay case by Azarian \etal
\cite{ElGamal_coop}. In \cite{ElGamal_coop}, it is shown that the NAF
scheme outperforms all previously proposed AF schemes in terms of the
fundamental diversity-multiplexing tradeoff~(DMT)\cite{Zheng_Tse} and
that it is optimal within the class of AF schemes in the single-relay
case. The superiority of the NAF scheme comes from the fact that the
source terminal is allowed to transmit during all the time, which
boosts up the multiplexing gain. However, even though they showed that
the DMT of this scheme can be achieved using a Gaussian random code of
sufficiently large block length, no practical coding scheme that
achieves the tradeoff has been proposed since then.

The main contributions of our work are summarized as follows:
\begin{enumerate} 
\item We extend the single-antenna NAF scheme proposed in
  \cite{Nabar,ElGamal_coop} to the multiple-antenna case. We establish
  a lower bound on the optimal DMT of the MIMO NAF channel. In
  particular, we show that the maximum diversity order of a
  single-relay MIMO NAF channel is lower-bounded by the sum of the
  maximum diversity order of the source-destination channel and the
  maximum diversity orders of the source-relay-destination product
  channels. This lower bound is tight when the source, relay and
  destination antenna number $\ns$, $\nr$ and $\nd$ satisfy
\[ \Abs{\ns-\nd} \geq \nr-1. \] 
\item We provide an explicit algebraic construction of \emph{short}
  block codes that achieves the optimal DMT of the general
  multiple-antenna multiple-relay NAF cooperative diversity scheme.
  Our algebraic code construction is inspired by the non-vanishing
  determinant~(NVD) space-time codes design for MIMO Rayleigh
  channels~\cite{Elia}. First, we show that for any linear fading
  Gaussian channel~(not only the Rayleigh channel as in
  \cite{Zheng_Tse,Elia})
\begin{equation}\label{eq:linear_model}
\my = \sqrt{\SNR}\,\mH \mx + \mz,
\end{equation}%
in the high SNR regime, the error event of a ``good'' space-time code
$\Xcal$~(which will be properly defined later) occurs only when the
channel is in outage. Therefore, the optimal DMT can always be
achieved by $\Xcal$. Then, as in \cite{ElGamal_coop}, we derive
equivalent signal models of the AF cooperative schemes in the
form~\Eq{eq:linear_model}, subject to certain input constraint~(\eg,
block diagonal for multiple-relay channel). Since codes that achieve
the optimal DMT of the equivalent channel~\Eq{eq:linear_model} also
achieve the optimal DMT of the corresponding cooperative channel,
optimal codes for an AF cooperative channel can be obtained from the
NVD criterion. As a result, we show that for a single-relay AF channel
with $\ns$ antennas at the source terminal, a $2\ns\times2\ns$ full
rate NVD space-time code~(\eg, the Golden code~\cite{Belfiore_golden}
for the single-antenna case and the $4\times4$ Perfect
code~\cite{Oggier-1} for the two-antenna case) can be directly applied
to construct an optimal block code. In the $N$-relay case, the optimal
code is constructed from a block-diagonal NVD space-time code with $N$
blocks. The performance of our construction is confirmed by simulation
results.
\end{enumerate}
 
The rest of the paper is outlined as follows. Section~\ref{sec:model}
introduces the system model and recalls the single-antenna NAF
protocol as well as the equivalent channel model. In
section~\ref{sec:mimo_naf}, we extend the NAF scheme to the MIMO
channel and develop a lower bound on the optimal
diversity-multiplexing tradeoff. The codes design criteria are derived
in section~\ref{sec:criteria} and the explicit algebraic construction
that satisfies the design criteria is provided in
section~\ref{sec:construction}. Section~\ref{sec:some-examples} shows
some examples of channel configuration and the parameters of the
corresponding optimal codes. Simulation results on our construction
are available in section~\ref{sec:results}.
Section~\ref{sec:conclusion} contains some concluding remarks. For
continuity of demonstration, most proofs are left in the appendices.

\section{System Model and Problem Formulation}
\label{sec:model}
\subsection{Channel Model}
\begin{figure}[!htbp]
\begin{center}
\includegraphics[angle=0,width=0.4\textwidth]{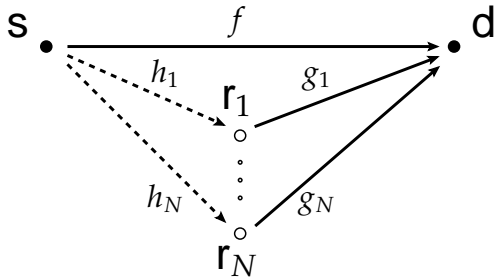}
\caption{A relay channel with one source~($\textsf{s}$), one destination~($\textsf{d}$)
 and  $N$ relays~($\textsf{r}_1,\ldots,\textsf{r}_N$).}
\label{fig:relay_model}  
\end{center}
\end{figure}

We consider a wireless network with $N+1$ sources~(users) and only one
destination. The channels are slow fading (or delay-limited), \ie, the
channel coherence time is much larger than the maximum delay that can
be tolerated by the application. For the moment, we assume that all
the terminals are equipped with only one antenna. The multi-antenna
case will be treated separately in section~\ref{sec:mimo_naf}. The
channel is shared in a TDMA manner, \ie, each user is allocated a time
slot for the transmission of its own data. Within the same time slot,
any of the other $N$ users can help the current user transmit its
information.  The extension to a more general orthogonal access scheme
is straightforward. Suppose that the network configuration is
symmetric.  Without loss of generality, we consider only one time slot
and the channel model becomes a single-user relay channel with one
source, $N$ relays and one destination, as shown in
\Fig{fig:relay_model}. Here, we exclude the multi-user case, where
information of more than one user can circulate at the same time in
the network~(\eg, the CMA-NAF scheme proposed in \cite{ElGamal_coop}).

In \Fig{fig:relay_model}, variables $f,h_i$ and $g_i, i=1,\ldots,N$
stand for the channel coefficients that remain constant during a block
of length $L$. As in the previous works we cite here, we assume that
all the terminals work in half duplex mode, \ie, they cannot receive
and transmit at the same time. The channel state information~(CSI) is
supposed to be known to the receiver but not to the transmitter.

\subsection{The Non-Orthogonal Amplify-and-Forward Relay Channel}
In our work, we consider the NAF protocol~(\cite{Nabar,
  ElGamal_coop}). In this scheme, the relays simply scale and forward
the received signal. However, unlike the orthogonal AF protocols, the
source can keep transmitting during the transmission of the relays.

\subsubsection{The Single-Relay Case}
\label{sec:single-relay-case}

\begin{figure}[!htbp]
\begin{center}
\includegraphics[angle=0,width=0.5\textwidth]{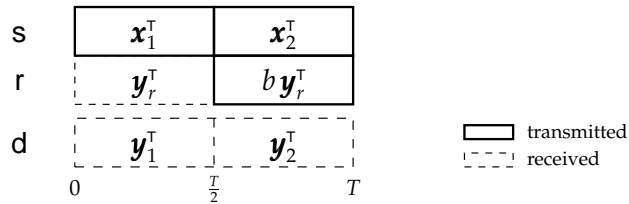}
\caption{The NAF frame structure of a single-relay channel, $b$ is the normalization factor such that $b\,{\my}_r$ is subject to the power constraint.}
\label{fig:frame_struct}  
\end{center}
\end{figure}

In the single-relay case, each frame is composed of two partitions of
$T/2$ symbols\footnote{It is shown in \cite{ElGamal_coop} that giving
  the same length to the two partitions is optimal in terms of the
  DMT.}.  The frame length $T$ is supposed to be smaller than the
channel coherence time $L$, \ie, the channel is static during the
transmission of a frame. The half duplex constraint imposes that the
relay can only transmit in the second partition. The frame structure
is illustrated in \Fig{fig:frame_struct}, from which we get the
following signal model
\begin{equation}
  \label{eq:signal_model}
  \setlength{\nulldelimiterspace}{0pt}
  \left\{
      \begin{aligned}
        \my_1 &= \sqrt{\pi_1\SNR}\,f\,\mx_1 + \mv_1 \\
        \my_r &= \sqrt{\pi_1\rho\,\SNR}\,h\,\mx_1 + \mw \\
        \my_2 &= \sqrt{\pi_3\SNR}\,g\,(b\,\my_r) +
        \sqrt{\pi_2\SNR}\,f\,\mx_2 + \mv_2
      \end{aligned}\right.
\end{equation}%
where $\mx_i,\my_i\in\CC^{T/2}$, $i=1,2$ are the transmitted signals
from the source with normalized power and the received signals at the
destination, respectively, in the $i^\th$ partition;
$\my_r\in\CC^{T/2}$ is the received signal at the relay in the first
partition; $\mv_1,\mv_2,\mw \in\CC^{T/2}$ are independent additive
white Gaussian noise~(AWGN) vectors with \iid unit variance entries;
the channels between different nodes are independently Rayleigh
distributed, \ie, $f, g, h\sim\CN[1]$; $\rho$ is the geometric gain
representing the ratio between the path loss of the source-relay link
and the source-destination link; $b$ is the normalization factor
satisfying $\EE\left\{\Norm{b\,\my_r}^2\right\}\leq\frac{T}{2}$, \ie,
\begin{equation*}
  \Abssqr{b} \leq \frac{1}{\pi_1\rho\,\SNR \Abssqr{h} + 1}.
\end{equation*}%
We consider a short term power constraint, \ie, the power allocation
factors $\pi_i$'s do not depend on the instantaneous channel
realization $f, g$ and $h$, but can depend on $\rho$ and $\SNR$. We
impose that $\sum_i\pi_i = 2$ so that $\SNR$ denotes the average
received SNR at the destination\footnote{The total transmit power in
  two partitions are $(\pi_1+\pi_2+\pi_3)\SNR$. Since the channel
  coefficients and the AWGN are normalized, $(\pi_1+\pi_2+\pi_3)\SNR$
  represents the average received SNR per two partitions as well.}.

As shown in \cite{ElGamal_coop}, the channel model
\Eq{eq:signal_model} is equivalent to $T/2$ channel uses of a $2\times2$
channel~:
\begin{equation*}
  \myt_i = \matrix{\sqrt{\pi_1\SNR}\,f & 0 \\\sqrt{\pi_1\pi_3\,\rho}\,\SNR\,b\,h\,g & \sqrt{\pi_2\SNR}\,f} \mxt_i + \matrix{0\\ \sqrt{\pi_3\SNR}\,b\,g}w_i +\mvt_i\quad\textrm{for}\ i=1,\ldots,T/2
\end{equation*}%
where $\tilde{\muu}_i = \trans{\bigl[\muu_1[i]\ \muu_2[i]\bigr]}$ for
$\muu\in\left\{\mx,\my,\mv\right\}$ and $\muu_k[i]$ denotes the
$i^\th$ symbol in the $k^\th$ partition. In the following, we consider
a more convenient normalized model
\begin{equation*}
  \myt_i = \sqrt{\SNR}\,\mHt \mxt_i + \mz_i\quad \for \ i=1,\ldots,T/2
\end{equation*}%
where $\mz_i\sim\CN$ is the equivalent AWGN and 
\begin{equation}
  \mHt \defeq \matrix{\sqrt{\pi_1} f & 0 \\\sqrt{\frac{\pi_1\pi_3\rho\SNR}{1+\pi_3\SNR\Abssqr{bg}}}b h g & \sqrt{\frac{\pi_2}{{1+\pi_3\SNR\Abssqr{bg}}}}f}.\label{eq:channelH}
\end{equation}%

\subsubsection{The Multiple-Relay Case}
\label{sec:multi-relay-case}

\begin{figure*}
\begin{center}
  \includegraphics[angle=0,width=0.7\textwidth]{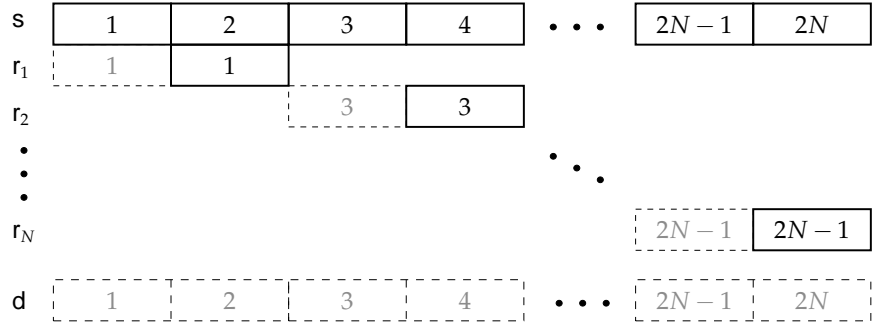}
\caption{The NAF frame structure of an $N$-relay channel. Solid box for transmitted 
signal and dashed box for received signal.}
\label{fig:frame_struct_nrelays}  
\end{center}
\end{figure*}

In the multiple-relay case, a superframe of $N$ consecutive
cooperation frames is defined, as shown in
\Fig{fig:frame_struct_nrelays}. It is assumed that the channel is
static during the transmission of the whole superframe~(of $NT$
symbols). The $N$ relays take turns to cooperate with the source.
Within each cooperation frame, the cooperation is in exactly the same
manner as in the single-relay case.  However, by allowing an encoding
over the whole superframe, a diversity order of $N+1$ is achieved.

\subsection{Diversity-Multiplexing tradeoff~(DMT)}
\begin{definition}[Multiplexing and diversity gain\cite{Zheng_Tse}]
  A coding scheme $\{\Ccal(\SNR)\}$ is said to achieve
  \emph{multiplexing gain} $r$ and \emph{diversity gain} $d$ if
\begin{equation*}
  \lim_{\SNR\to\infty} \frac{R(\SNR)}{\log\SNR} = r \quad
\textrm{and}\quad 
  \lim_{\SNR\to\infty} \frac{\log\Pe(\SNR)}{\log\SNR} = -d
\end{equation*}
where $R(\SNR)$ is the data rate measured by bits per channel use~(PCU)
and $\Pe(\SNR)$ is the average error probability using the maximum
likelihood~(ML) decoder.
\end{definition}

The optimal DMT of the single-antenna $N$-relay NAF channel
\begin{equation}
  d_{\NAF}(r) = (1-r)^+ + N(1-2r)^+\label{eq:DMT}
\end{equation}%
is found in \cite{ElGamal_coop}, where the achievability is proved by
using a Gaussian random code with a sufficiently long block length.

\section{The NAF Scheme for MIMO Channel}
\label{sec:mimo_naf}
In this section, we generalize the NAF protocol to the MIMO case,
where each terminal is equipped with multiple antennas. The
notation~$(\ns, \nr, \nd)$ will be used to denote a single-relay
channel with $\ns,\nr$ and $\nd$ antennas at the source, relay and
destination. All matrix variables defined for a single-relay channel
apply to a multiple-relay channel with an index $i$ denoting the
$i^\th$ relay.

\subsection{Signal Model}
\label{sec:signal-model}
For convenience of demonstration, we only present the signal model of
the single-relay channel. The extension to the multiple-relay case is
straightforward.
\subsubsection{$\nr\leq \ns$}
\label{sec:n_rn_s}
In the case $\nr\leq \ns$, a direct generalization of
\Eq{eq:signal_model} is as follows~:
\begin{equation}
  \label{eq:signal_model_MIMO}
  \setlength{\nulldelimiterspace}{0pt}
  \left\{
      \begin{aligned}
        \mY_1 &= \sqrt{\pi_1\SNR}\,\mF\mX_1 + \mV_1 \\
        \mY_r &= \sqrt{\pi_1\rho\,\SNR}\,\mH\mX_1 + \mW \\
        \mY_2 &= \sqrt{\pi_3\SNR}\,\mG(\mB\mY_r) +
        \sqrt{\pi_2\SNR}\,\mF\mX_2 + \mV_2
      \end{aligned}\right.
\end{equation}%
where $\mF,\mG,\mH$ are $\nd\times \ns, \nd\times \nr$ and $\nr\times
\ns$ independent matrices, respectively, with zero mean unit variance
\iid Gaussian entries; $\mX_i$'s are $\ns\times \frac{T}{2}$ matrices
with \iid zero mean unit variance entries, representing the space-time
signal from the source; $\mV_1,\mV_2$ and $\mW$ are independent AWGN
matrices with normalized \iid entries; the power allocation factors
$\pi_i$'s satisfy $\ns(\pi_1+\pi_2)+\nr \pi_3=2$ so that $\SNR$
denotes the received SNR per receive antenna at the destination; $\mB$
is an $\nr\times \nr$ matrix equivalent to the ``normalization
factor'' $b$ in the single-antenna case and is subject to the power
constraint $\EE\Bigl\{\Frob{\mB\mY_r}\Bigr\} \leq \frac{T}{2}\nr$
which can be simplified to
\begin{equation}\label{eq:constraintB}
  \Tr\Bigl\{\left(\Id+\pi_1\rho\SNR\mH\transc{\mH}\right)\transc{\mB}\mB\Bigr\} \leq \nr.
\end{equation}%
Now, as in the single-antenna case, we obtain an equivalent
single-user MIMO channel
\begin{equation}
  \myt_i = \sqrt{\SNR}\He\mxt_i + \mz_i \quad \for \ i=1,\ldots,T/2\label{eq:channel_MIMO}
\end{equation}%
where $\mxt_i = \trans{\bigl[\mX_1[i]^\ssT \ \mX_2[i]^\ssT \bigr]}$
and $\myt_i = \trans{\bigl[\mY_1[i]^\ssT \ \mY_2[i]^\ssT \bigr]}$ are
the vectorized transmitted and received signals with $\mM[i]$ denoting
the $i^\th$ column of the matrix $\mM$; $\mz_i\sim\CN[\Id]$ is the
equivalent AWGN; the equivalent channel matrix $\He$ is
\begin{equation}
  \He \defeq \matrix{\sqrt{\pi_1} \mF & \mbs{0} \\\sqrt{\pi_1\pi_3}\,\mSigmasqrt\mP\mH & \sqrt{\pi_2}\,\mSigmasqrt\mF}\label{eq:channelH_MIMO}
\end{equation}%
with
\begin{equation}
  \mP \defeq \sqrt{\SNR}\, \mG\mB  \label{eq:def_P}
\end{equation}%
and $\mSigmasqrt$ being the whitening matrix satisfying
$\left(\mSigmasqrttrc\,\mSigmasqrt\right)^{-1} =
\left(\mSigmasqrt\mSigmasqrttrc\right)^{-1} = \Id +
\pi_3\mP\transc{\mP} \defeq \mSigma$.


\subsubsection{$\nr > \ns$}
\label{sec:n_r-geq-n_s}
With more antennas at the relay than at the source, the relay can do
better than simple forwarding. In this case, the received signal at
the relay is in the $\ns$-dimensional subspace generated by the $\ns$
eigenmodes of $\mH$, represented by $\mU_\mH$ from the singular
value decomposition $\mH = \mU_\mH \mSigma_{\mH} \transc{\mV}_\mH$.
Since the relay-destination channel $\mG$ is isotropic, it is of no
use to forward the received signal in more than $\nr$
antennas~(spatial directions).

However, by using only a subset of the antennas at the relay, we
cannot obtain all the diversity gain provided by the channel $\mG$. To
exploit all the available diversity, we propose two schemes.  The
first scheme is the virtual $N$-relay scheme. Since no CSI is
available at the transmitter and therefore no antenna combination is
{\it a priori} better than the others, one solution is to use all the
$\binom{\nr}{\ns}$ antenna combinations equally. Intuitively, the
typical outage event is that all the antenna combinations are in deep
fade, which implies that the channel $\mG$ is also in outage. In this
scheme, a superframe of $\binom{\nr}{\ns}$ cooperation frames is
constructed.  Within each cooperation frame, a different combination
of $\ns$ antennas is used.  The second scheme is the antenna selection
scheme, which is used only when limited feedback from the destination
to the relay is available. In this case, the destination tells the
relay which antenna combination is optimal according to a given
criterion~(\eg,  maximization of mutual information). With this
scheme, maximum diversity gain is obtained without a superframe
structure, which means a significant reduction of coding-decoding
complexity.

As an example, let us consider a $(1,\nr,1)$ relay channel with
$\nr>1$. In this case, the received signal at the relay can be
projected into a one-dimensional subspace. Consider a superframe of
$\nr$ cooperation frames. In each cooperation frame, a different relay
antenna is used.  This scheme is virtually an $N$-relay single-antenna
channel. The only difference is that the equivalent source-relay link,
which is $\sqrt{\sum_i\Abssqr{h_i}}$ after the matched filter
operation, is the same for the $N$ virtual relays. In this scheme, the
achievable diversity order of the source-relay-destination link is
$\nr$, since the channel is in outage only when
$\sqrt{\sum_i\Abssqr{h_i}}$ is in deep fade or all the $\nr$
relay-destination links are in deep fade. When feedback is possible,
the antenna selection scheme can be used. The difference from the
first scheme is that only the antenna with maximum relay-destination
channel gain~(say, $\Abs{g_{\max}}\defeq \D\max_{i=1\ldots
  \nr}\left\{\Abs{g_i}\right\}$) is used in the relaying phase.
Therefore, only one cooperation frame is needed and the diversity
order is also $\nr$.


\subsection{Optimal  Diversity-Multiplexing Tradeoff: a Lower Bound}
\label{sec:divers-mult-trad}
With the discussion above, considering the case $\nr\leq \ns$ is
without loss of generality. In addition, since the destination is
usually equipped with more antennas than the relays are in practice,
we will restrict ourselves to the case $\nd\geq \nr$ hereafter.  In
the rest of this section, we study the optimal DMT of the MIMO NAF
cooperative channel. Unlike the single-antenna case, a closed form
expression of the DMT of the MIMO NAF channel is difficult to obtain,
since the probability distribution~(in the high SNR regime or not) of
the eigenvalues of $\He$ defined in \Eq{eq:channelH_MIMO} is unknown.
In the following, we will derive a lower bound on the tradeoff, as a
generalization of the DMT of the single-antenna NAF channel provided in
\cite{ElGamal_coop}. To this end, we first study the DMT of a Rayleigh
product channel.

\subsubsection{DMT of a Rayleigh Product Channel}
\newcommand{\dAl}{\underline{d}_{\mA}}
\newcommand{\dAu}{\bar{d}_{\mA}} \newcommand{\dA}{d^*_{\mA}}
\begin{proposition}\label{prop:prod}
  Let $\mG,\mH$ be $n\times l, l\times m$ independent matrices with
  \iid entries distributed as $\CN[1]$. Assume that $m\geq l, n\geq l$
  and define $\mA\defeq \mG\mH, \Delta\defeq \Abs{m-n},
  q\defeq\min\left\{m,n\right\}$, then the optimal DMT curve
  $\dA(r)$ of the Rayleigh product channel, \ie, the channel defined
  by
  \begin{equation}
    \my = \sqrt{\frac{\SNR}{l\cdot m}}\mA\mx + \mz \label{eq:H1H2}
  \end{equation}%
  with $\mz\sim\CN[1]$ being the AWGN is a piecewise-linear function
  connecting the points $(s,\dA(s)),s=0,\ldots,l$, where
  \begin{equation}
    \label{eq:dmt_pr}
    \dA(s) = (l-s)(q-s) - \frac{1}{2}\left\lfloor \frac{\left[(l-\Delta-s)^+\right]^2}{2}\right\rfloor.
  \end{equation}

\end{proposition}
\begin{proof}
  See \App{app:proof_prop1}. A more general result is given by
  \cite{SY_JCB_ds}, where a Rayleigh product channel is seen as a
  special case of the double scattering channels and the assumption
  $m\geq l,n\geq l$ is unnecessary.
\end{proof}
\begin{remark}
  From \Prop{prop:prod}, we note that
  \begin{enumerate}[(i)]
  \item $\dA(r)$ only depends on $\Delta$ and $q$, which means that
    interchanging $m$ and $n$ does not change $\dA(r)$, which is
    obvious if we consider the fact that $\mA\transc{\mA}$ has the same
    eigenvalues as $\transc{\mA}\mA$;
  \item $\dA(s)$ is upper-bounded by
    $\dAu(s)\defeq\min\bigl\{d_{\mH}(s),d_{\mG}(s)\bigr\}=(l-s)(q-s)$
    and coincides with it when $s\geq l-\Delta-1$;
  \item when $\Delta\geq l-1$, $\dA(s)=\dAu(s), \forall s$.  Without
    loss of generality, assume that $n\geq m$. Intuitively, by
    increasing $n$ and keeping $m$ unchanged, the diversity gain of
    the channel $\mG$ is increasing and so is that of the product
    channel $\mA$. When $n-m=l-1$, the best possible DMT for $q=m$ is
    achieved.  With this $n$, the fading effect of $\mG$ vanishes,
    since $\dA(s)=d_{\mH}(s)$. In other words, it is of no use to have
    $n>l+m-1$, for $m$ and $l$ fixed.

  \end{enumerate}
\end{remark}

\begin{figure*}[ht]
  \begin{center}
\includegraphics[angle=0,width=12cm]{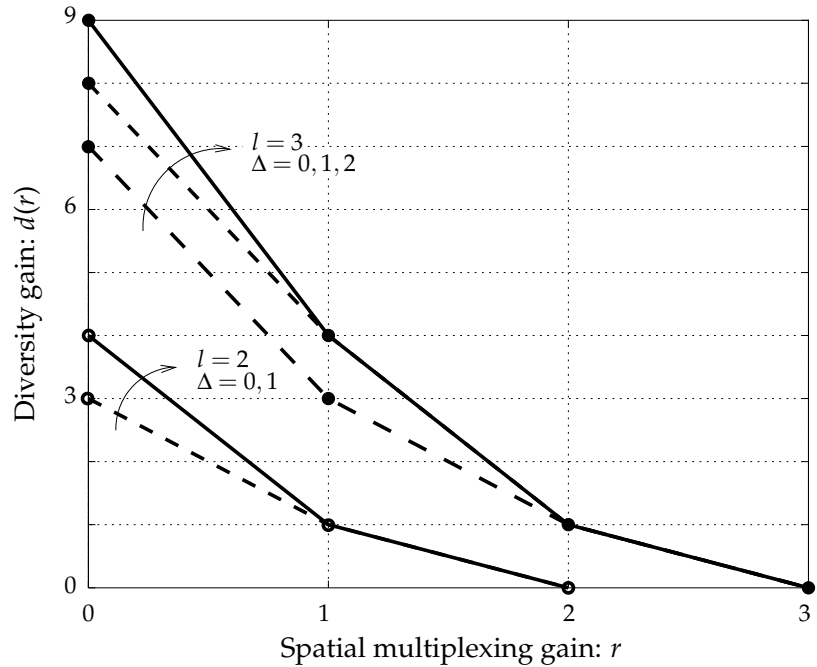}
\caption{Optimal DMT: Rayleigh \vs Rayleigh product channel. $\min\{m,n\}=l=2,3$. 
$\Delta = 0,\ldots,l-1$.}
\label{fig:dmt_pr}    
  \end{center}
\end{figure*}

The optimal DMTs of a Rayleigh product channel for $l=2,3$
are illustrated in \Fig{fig:dmt_pr}. As indicated in remark~1, when
$\Delta^*=1,2$ for $l=2,3$, the tradeoffs of the Rayleigh product
channels are the same as those of the corresponding Rayleigh channels,
\ie, the $2\times2$ and $3\times3$ Rayleigh channels.

\NC{\detexp}{\det\left[\exp{\left(-\SNR^{-(\alpha_j-\beta_i)}\right)}\right]}

\subsubsection{DMT of a MIMO NAF Channel}

\begin{theorem}\label{thm:dmt_MIMONAF}
  For a single-relay MIMO NAF channel \Eq{eq:signal_model_MIMO}, we
  have
  \begin{equation} 
    d_{\NAF}(r) \geq d_{\mF}(r) + d_{\mG\mH}(2r) \label{eq:dNAFMIMO}
  \end{equation}%
  where $d_{\mF}(r)$ and $d_{\mG\mH}(r)$ are the optimal DMT of the
  fading channel $\mF$ and the fading product channel $\mG\mH$,
  respectively.
\end{theorem}
\begin{proof}
   See \App{app:proof_thm1}.
\end{proof}

\begin{figure*}[ht]
  \begin{center}
    \includegraphics[angle=0,width=12cm]{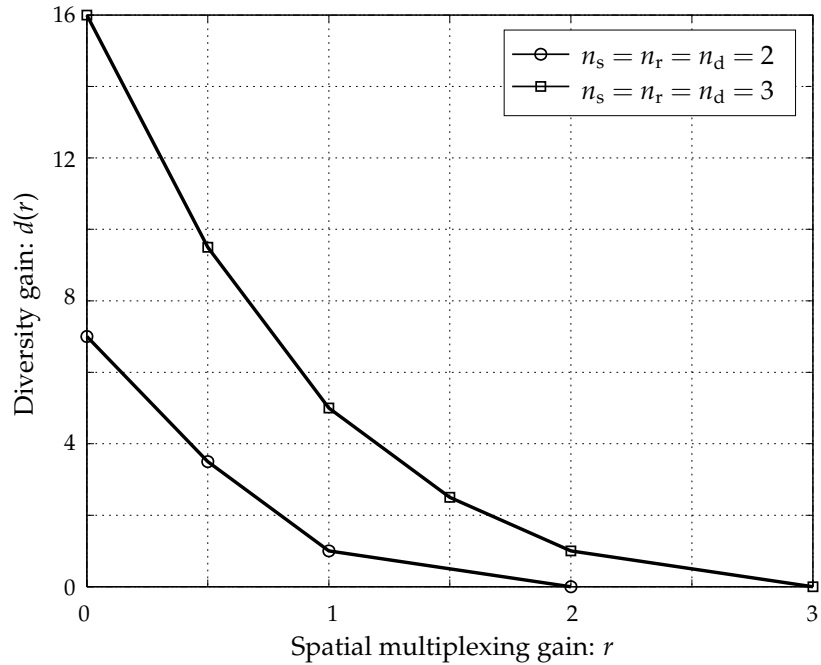}
\caption{Lower bound on the optimal DMT: $\ns=\nr=\nd=2,3$.}
\label{fig:naf_mimo_dmt}    
  \end{center}
\end{figure*}

The interpretation of \Eq{eq:dNAFMIMO} is as follows. First, since the
transmitted signal passes through the source-destination link all the
time, a diversity gain $d_{\mF}(r)$ can be obtained. Then, due to the
half duplex constraint, only half of the transmitted signal is
protected by the source-relay-destination link, \ie, the channel
defined by $\mG\mH$.  Fig.~{\ref{fig:naf_mimo_dmt}} shows the lower
bound~\Eq{eq:dNAFMIMO} for a Rayleigh channel with $\ns=\nr=\nd=2$ and
$3$.

This result of theorem~\ref{thm:dmt_MIMONAF} can be generalized to the
$N$-relay case. Let $\mG_i$ and $\mH_i$ be channels related to the
$i^\th$ relay and be similarly defined as $\mG$ and $\mH$. The
following theorem gives a lower bound on the optimal DMT of the
$N$-relay MIMO NAF channel.
\begin{theorem}\label{thm:dmt_MIMONAF_N}
  For an $N$-relay MIMO NAF channel, we have
  \begin{equation*}
    d_{\NAF}(r) \geq d_{\mF}(r) + \min_{\mtheta:\sum_{i} \theta_i=1}\sum_{i=1}^{N}d_{\mG_i\mH_i}(2N\theta_i r).
  \end{equation*}%
  In particular, when all the relays have the same number of antennas,
  we have
    \begin{equation}\label{eq:lb3}
    d_{\NAF}(r) \geq d_{\mF}(r) + N d_{\mG\mH}(2r).
  \end{equation}%
\end{theorem}

\begin{proof}
   See \App{app:proof_thm2}.
\end{proof}
From \Eq{eq:DMT}, we see that the bound in \Eq{eq:lb3} is actually the
optimal tradeoff in the single-antenna case~(\ie, $\ns=\nr=\nd=1$).

\begin{corollary} \label{coro:1}
  Let $d_i\defeq \min\bigl\{d_{\mG_i}(0),d_{\mH_i}(0)\bigr\},
  i=1,\ldots, N$. Then, we have
  \begin{equation}
    d_{\mF}(0) + \sum_{i=1}^{N} d_{i}(0) \geq d_{\NAF}(0) \geq d_{\mF}(0) + \sum_{i=1}^{N}d_{\mG_i\mH_i}(0).\label{eq:max_d}
  \end{equation}%
  If all the channels are Rayleigh distributed and $\Abs{m-n}\geq
  l_i-1,\forall i$, then, we have
  \begin{equation}
    d_{\NAF}(0) = d_{\mF}(0) + \sum_{i=1}^{N}d_{i}(0).\label{eq:eq}
  \end{equation}%
\end{corollary}
\newcommand{\Hc}{\hat{\mH}} 
\newcommand{\Hctrc}{\transc{\hat{\mH}}}
\newcommand{\Hetrc}{\transc{\He}}
\begin{proof}
  See \App{app:proof_thm2}.
\end{proof}

\section{Optimal Codes Design Criteria}
\label{sec:criteria}
In this section, we will derive design criteria for a family of
\emph{short} codes to achieve the optimal DMT of an $N$-relay MIMO NAF
channel.

\subsection{A General Result}
Let us first define the ``good'' code mentioned in
section~\ref{sec:intro}.
\begin{definition}[Rate-$n$ NVD code]
  Let $\Acal$ be an alphabet that is scalably dense, \ie, for $0\leq
  r\leq n$
\begin{eqnarray*}
\Abs{\Acal(\SNR)} \asympteq \SNR^{\frac{r}{n}} \quad \textrm{and} \\
a\in\Acal(\SNR) \Rightarrow \Abssqr{a} \asymptleq \SNR^{\frac{r}{n}} 
\end{eqnarray*}%
Then, an $\nT\times \nT$ space-time code $\Xcal$ is called a rate-$n$
NVD code if it
\begin{enumerate}
\item is $\Acal$-linear\footnote{$\Xcal$ is $\Acal$-linear means that
    each entry of any codeword $\mX\in\Xcal$ is a linear combination
    of symbols from $\Acal$.};
\item transmits on average $n$ symbols PCU from the signal
  constellation $\Acal$;
\item has the non-vanishing determinant~(NVD) property\footnote{NVD
    means that $\Abs{\det(\mX_i-\mX_j)}\geq\kappa>0,\ \forall
    \mX_i,\mX_j\in\Xcal, \mX_i\neq\mX_j$ with $\kappa$ a constant
    independent of the SNR.}.
\end{enumerate}
\end{definition}

The following theorem is fundamental to our construction.
\begin{theorem} \label{thm:dmt}
For any linear block fading channel 
\begin{equation*}
  \my = \sqrt{\SNR}\mH\mx + \mz 
\end{equation*}%
where $\mH$ is an $\nR\times \nT$ channel and $\mz\sim\CN$ is the
AWGN, the achievable DMT of a rate-$n$ NVD code $\Xcal$ satisfies
\begin{equation}
  \label{eq:dmt_lowerbound}
  \dX(r) \geq \dout\left(\frac{q}{n}r\right)
\end{equation}%
where $q\defeq\min\{\nR,\nT\}$ and $\dout(r)$ is the outage upper
bound of the DMT for the channel $\mH$.
\end{theorem}
\begin{proof}
  See \App{app:proof_thm3}.
\end{proof}

In particular, for a full rate code ($n = q$), the upper bound
$\dout(r)$ is achievable. This theorem implies that the NVD property
is fundamental for $\Xcal$ to achieve all the diversity gain $d$, for
any linear fading channel. For a given diversity gain $d$, the
achievable multiplexing gain $r$ of such $\Xcal$ is a shrunk version
of $r_{\out}(d)$, the best that we can have for channel $\mH$.

One of the consequences of \Thm{thm:dmt} is the possibility of
constructing optimal codes~(in terms of the DMT) based on the NVD
criteria for some channels. For example, we can get an equivalent MIMO
space-time model for the single-antenna fast fading channel~(also
called a Gaussian parallel channel) as
\begin{equation}
  \label{eq:ff}
  \mY = \diag\left(h_1,\ldots,h_N\right)\mX + \mZ
\end{equation}%
with $\mX,\mY$ and $\mZ$ diagonal $N\times N$ matrices. The best code
that we can have is a rate-$1$ NVD code $\Xcal$ due to the diagonal
constraint. According to \Thm{thm:dmt}, we have $\dX(r)\geq\dout(Nr)$
with $\dout(r)$ the DMT of \Eq{eq:ff} without the diagonal constraint.
In fact, we can verify that $\dout(Nr)$ coincides with the DMT of the
fast fading channel. The NVD criterion includes the product distance
criterion since the determinant of a diagonal matrix is the product of
the diagonal entries. In addition, it implies that the product
distance should be non-vanishing as the constellation size increases.

Note that another such general result as theorem~\ref{thm:dmt}, has
been derived independently in \cite{Tavildar}. In \cite{Elia}, the NVD
property is derived from the mismatched eigenvalue bound~(worst case
rotation) while the results in \cite{Tavildar} are derived using the
worst case codeword error probability, which is effectively the same
thing as the worst case rotation. Theorem~\ref{thm:dmt} is a
generalization of the result in \cite{Elia}~(for the full rate codes)
to a rate-$n$ code. This result is more adapted to the algebraic
construction of explicit codes for the relay channel.

\subsection{Design Criteria}
With \Thm{thm:dmt}, we are ready to give out the design criteria of
the optimal codes for the NAF cooperative channel. The following
theorem states the main result of our work.
\begin{theorem}\label{thm:achievability}
  Let $\Xcal$ be a rate-($2\ns$) NVD block diagonal code, \ie,
  \begin{equation*}
\mX=\begin{bmatrix} \mXi_{1} & \cdots & 0\\
\vdots & \ddots & \vdots\\
0 & \cdots & \mXi_{N}\end{bmatrix},\quad\forall\mX\in\Xcal
  \end{equation*}%
  where $\mXi_i$'s are $2\ns\times2\ns$ matrices. Now consider an
  equivalent code $\Ccal$ whose codewords are in the form
  \begin{equation*}
    \mC = \matrix{\mC_1&\ldots&\mC_N}
  \end{equation*}%
  with
  \begin{equation*}
    \mC_i \defeq \matrix{\mXi_i\left(1\!:\!\ns, 1\!:\!2\ns\right)&\mXi_i\left(\ns\!+\!1\!:\!2\ns, 1\!:\!2\ns\right)}.
  \end{equation*}%
  Then, $\Ccal$ achieves the optimal DMT of the $N$-relay MIMO NAF
  channel with $\ns$ transmit antennas at the source, by transmitting
  $\mC_i$ in the $i^\th$ cooperation frame. The code $\Ccal$ is of
  length $4N\ns$.
\end{theorem}

\begin{proof}
  See \App{app:proof_thm4}.
\end{proof}
In section~\ref{sec:mimo_naf}, a lower bound on the optimal DMT of a
MIMO NAF channel is derived. Here, theorem~\ref{thm:achievability}
shows that the \emph{exact} optimal tradeoff can always be achieved by
a code $\Ccal$, even though we cannot obtain its closed form
expression.

\section{A Unified Construction Framework}
\label{sec:construction}
\subsection{Notations and Assumptions}

We assume that the modulation used by the source is either a QAM or a
HEX modulation. The fields representing the modulated symbols will be
either $\mathbb{Q}(i)$ or $\mathbb{Q}(j)$. We denote it as
$\mathbb{P}$. For each algebraic number field $\KK$, the ring of
integers is denoted $\OK$.

\subsection{Behavior of the Codewords\label{sub:criteria}}

We recall that a codeword $\mX$ is represented by a block
diagonal matrix
\begin{equation}
\mX=\begin{bmatrix} \mXi_{1} & \cdots & 0\\
\vdots & \ddots & \vdots\\
0 & \cdots & \mXi_{N}\end{bmatrix}\label{eq:codeword-structure}
\end{equation}
with $\mXi_{i},\,\, i=1,\ldots,N$ being a square $2\ns\times2\ns$ matrix.
The criteria to fulfill are the following~:
\begin{enumerate}
\item \emph{full rate}~: the number of QAM or HEX independent symbols
  in a codeword is equal to $N\cdot(2\ns)^{2}$ corresponding to a
  multiplexing gain of $\ns$ symbols PCU;
\item \emph{full rank}~: 
\begin{equation}
    \min_{{\mX_1,\mX_2\in\mathcal{C}}\atop{\mX_1\neq\mX_2}}\textrm{rank}(\mX_1-\mX_2)=N\cdot(2\ns);\label{eq:rank-criterion}\end{equation}
\item \emph{non-vanishing determinant}~:
\begin{equation}
\min_{{\mX_1,\mX_2\in\mathcal{C}}\atop{\mX_1\neq\mX_2}} \left|\det(\mX_1-\mX_2)\right|^{2}\geq\kappa\label{eq:non-vanishing-det}\end{equation}
with $\kappa$ being some strictly positive constant. 
\end{enumerate}

\subsection{Codes Construction}

We use the same methods as in \cite{Oggier-1}. Some particular cases
can be found in \cite{Belfiore-1,Belfiore_golden}. The main difference
is in the choice of the base field $\mathbb{F}$. In \cite{Oggier-1},
this base field was equal to $\mathbb{P}$. Here, we choose a Galois
extension of $\mathbb{P}$ with degree $N$ and denote $\tau_{i},\,\,
i=1,\ldots,N$ the elements of its Galois group
$\mathcal{G}al_{\mathbb{F}/\mathbb{P}}$.  Now, we construct a cyclic
algebra whose center is $\mathbb{F}$.  We need a cyclic extension over
$\mathbb{F}$ of degree $2\ns$. We denote it $\mathbb{K}$. The
generator of its Galois group is $\sigma$.  The code construction
needs two steps.

\begin{enumerate}
\item Construction of the cyclic algebra
\begin{equation}
\mathcal{A}=\left\{ \left.\sum_{i=0}^{2\ns-1}z_{i}\cdot e^{i}\ \right|z_{i}\in\mathbb{K}\right\} \label{eq:cyclic-algebra}
\end{equation}
such that $e^{2\ns}=\gamma\in\mathbb{F}$ and $z_{i}\cdot
e=e\cdot\sigma\left(z_{i}\right)$.  In the matrix representation, we
have
\begin{equation*}
  e=\begin{bmatrix}
  0 & 1 & 0 & \cdots & 0\\
  0 & 0 & 1 & \ddots & \vdots\\
  \vdots &  & \ddots & \ddots & 0\\
  0 & \vdots &  & \ddots & 1\\
  \gamma & 0 & \cdots &  & 0\end{bmatrix}
\end{equation*}
and
$z_{i}=\textrm{diag}\left(z_{i},\sigma\left(z_{i}\right),\sigma^{2}\left(z_{i}\right),\ldots,\sigma^{2\ns-1}\left(z_{i}\right)\right)$.
\item {Application of the embeddings of $\mathbb{F}/\mathbb{P}$.} 
\end{enumerate}
In terms of matrices, we construct, in step 1, the square
$2\ns\times2\ns$ matrix $\mXi$. Then, by applying the embeddings of
$\mathbb{F}/\mathbb{P}$, the codeword is
\begin{equation}
  \mX=\begin{bmatrix}
    \tau_{1}(\mXi) & \cdots & 0\\
    \vdots & \ddots & \vdots\\
    0 & \cdots &
    \tau_{N}(\mXi)\end{bmatrix}\label{eq:canonical-embedding}
\end{equation}
where we can identify $\tau_{i}(\mXi)=\mXi_{i}$ from
(\ref{eq:codeword-structure}).  As usual, we restrict the information
symbols to be in $\mathcal{O}_{\mathbb{P}}$, that is, $\mathbb{Z}[i]$
(QAM symbols) or $\mathbb{Z}[j]$ (HEX symbols).  So, instead of being
in $\mathbb{F}$, we will be in $\mathcal{O}_{\mathbb{F}}$ and in the
same way, we will be in $\mathcal{O}_{\mathbb{K}}$ instead of
$\mathbb{K}$. The infinite space-time code is defined as being the set
of all matrices \begin{equation} \mathcal{C}=\left\{
    \mX=\begin{bmatrix}
      \tau_{1}\left(\D\sum_{i=0}^{2\ns-1}z_{i}e^{i}\right) & \cdots & 0\\
      \vdots & \ddots & \vdots\\
      0 & \cdots &
      \tau_{N}\left(\D\sum_{i=0}^{2\ns-1}z_{i}e^{i}\right)\end{bmatrix}\right\}.
  \label{eq:space-time-code}\end{equation}

\subsection{Codes Properties}

\begin{lemma}
  The code $\mathcal{C}$ of (\ref{eq:space-time-code}) is full rate.
\end{lemma}
\begin{proof}
  In the submatrix $\mXi$, there are $2\ns$ independent elements of
  $\mathcal{O}_{\mathbb{K}}$.  Each element in
  $\mathcal{O}_{\mathbb{K}}$ is a linear combination of $2\ns$
  elements of $\mathcal{O}_{\mathbb{F}}$. Finally, each element of
  $\mathcal{O}_{\mathbb{F}}$ is a linear combination of $N$ QAM or HEX
  symbols. So, each codeword $\mX$ is a linear combination of
  $N\cdot(2\ns)^{2}$ QAM or HEX symbols.
\end{proof}
\begin{lemma}
  If $\gamma,\gamma^2,\ldots,\gamma^{2n_s-1}\notin
  N_{\mathbb{K}/\mathbb{F}}(\mathbb{K})$, then the code $\mathcal{C}$
  is full rank.
\end{lemma}
\begin{proof}
  In \cite{Pierce}, it is proved that if
  $\gamma,\gamma^2,\ldots,\gamma^{2n_s-1}\notin
  N_{\mathbb{K}/\mathbb{F}}(\mathbb{K})$, then the cyclic algebra
  $\mathcal{A}$ is a division algebra (each element has an inverse).
\end{proof}
\begin{lemma}
  If $\gamma,\gamma^2,\ldots,\gamma^{2n_s-1}\notin
  N_{\mathbb{K}/\mathbb{F}}(\mathbb{K})$, then the code $\mathcal{C}$
  has a non-vanishing determinant, more precisely
\begin{equation*}
\delta_{\min}\defeq\min_{{\mX\in\mathcal{C}}\atop{
\mX\neq0}} \left|\det\mX\right|^{2}\in\mathbb{Z}^+\!\setminus\!\{0\}\geq1.
\end{equation*}
\end{lemma}

\begin{proof}
  Because of the structure of $\mX$, its determinant is\[
  \det\mX=\prod_{i=1}^{N}\det\tau_{i}(\mXi)=\prod_{i=1}^{N}\tau_{i}\left(\det(\mXi)\right).\]
  But, $\det(\mXi)$ is the reduced norm of
  $\sum_{i=0}^{2\ns-1}z_{i}e^{i}$ thus it belongs to
  $\mathcal{O}_{\mathbb{F}}$. So, \[
  \prod_{i=1}^{N}\tau_{i}\left(\det(\mXi)\right)=N_{\mathbb{F}/\mathbb{P}}\left(\det(\mXi)\right)\in\mathcal{O}_{\mathbb{P}}\]
  with $\mathcal{O}_{\mathbb{P}}=\mathbb{Z}[i]$ or
  $\mathcal{O}_{\mathbb{P}}=\mathbb{Z}[j]$.  Since $\det(\mX)\neq0$
  unless $\mX=\mbs{0}$, we get $\delta_{\min}\geq1$.
\end{proof}
Finally, the following result is derived.
\begin{theorem}
  The code $\mathcal{C}$ of (\ref{eq:space-time-code}) with
  $z_{i}\in\mathcal{O}_{\mathbb{K}}$ or a subspace of
  $\mathcal{O}_{\mathbb{K}}$ (which will be in the following an ideal
  of $\mathcal{O}_{\mathbb{K}}$) achieves the DMT of the MIMO NAF
  cooperative channel when $N$ is the number of relays and $\ns$ is
  the number of antennas at the source.
\end{theorem}
\begin{proof}
  The proof is straightforward and uses the results of the $3$ above
  lemmas.
\end{proof}

\subsection{Shaping}

As in \cite{Belfiore_golden,Oggier-1}, we may be interested in
constructing codes that achieve the DMT and that behave well in terms
of error probability even for small alphabets such as QPSK (4QAM).  In
that case, we add another constraint to our codes design, the shaping
factor. This new constraint implies that $\left|\gamma\right|=1$.
Moreover, as in \cite{Belfiore_golden,Oggier-1}, the linear transform
that sends the vector composed by the $N\cdot{(2\ns)}^2$ QAM or HEX
information symbols to $\textrm{vec}(\mX)$ has to be unitary.  The
following examples will illustrate this claim.

\section{Some Examples}
\label{sec:some-examples}
We give some examples of the code construction. Our code for an
$N$-relay $k$-antenna channel is denoted $\Ccal_{N,k}$.

\subsection{The Golden Code \cite{Belfiore_golden} is Optimal for the  Single-Relay Single-Antenna 
NAF Channel}

In the case of single-relay single-antenna channel, the codewords are
$2\times2$ matrices. Because the Golden code satisfies to all the
criteria of subsection \ref{sub:criteria}, it achieves the optimal DMT
of the channel.

\subsection{Two Relays, Single Antenna}

Optimal codes for the case $N>1$ relays cannot be found in the literature.

For the $2$-relay case, we propose the following code. Codewords are
block diagonal matrices with $2$ blocks. Each block is a $2\times2$
matrix. Let $\mathbb{P}=\mathbb{Q}(i)$ and
$\mathbb{F}=\mathbb{Q}\left(\zeta_{8}\right)$ with
$\zeta_{8}=e^{\frac{i\pi}{4}}$ be an extension of $\mathbb{Q}(i)$ of
degree $2$. We choose
$\mathbb{K}=\mathbb{F}\left(\sqrt{5}\right)=\mathbb{Q}\left(\zeta_{8},\sqrt{5}\right)$.
In fact, we try to construct the Golden code on the base field
$\mathbb{Q}\left(\zeta_{8}\right)$ instead of the base field
$\mathbb{Q}(i)$. Moreover, the number $\gamma$ is no more equal to $i$
because $i$ is a norm in $\mathbb{Q}\left(\zeta_{8}\right)$
($i=N_{\mathbb{K}/\mathbb{F}}\left(\zeta_{8}\right)$). We choose here,
in order to preserve the shaping of the code, $\gamma=\zeta_{8}$.  We
prove in appendix \ref{zeta8-not-norm} that $\zeta_{8}\notin
N_{\mathbb{K}/\mathbb{F}}(\mathbb{K})$ and thus that this code
satisfies to the full rank and the NVD conditions. Such a code uses
$8$ QAM symbols. Let $\theta=\frac{1+\sqrt{5}}{2}$, $\gamma=\zeta_{8}$
and $\sigma:\theta\mapsto\bar{\theta}=\frac{1-\sqrt{5}}{2}$.  The ring
of integers of $\mathbb{K}$ is $\mathcal{O}_{\mathbb{K}}=\left\{
  a+b\theta\mid a,b\in\mathbb{Z}\left[\zeta_{8}\right]\right\} $.  Let
$\alpha=1+i-i\theta$ and $\bar{\alpha}=1+i-i\bar{\theta}$.  Codewords
are given by
\[
\mX=\begin{bmatrix}
  \mXi & \mbs{0}\\
  \mbs{0} & \tau(\mXi)\end{bmatrix}\] with \[ \mXi=\frac{1}{\sqrt{5}}
\begin{bmatrix}
  \alpha\cdot\left(s_{1}+s_{2}\zeta_{8}+s_{3}\theta+s_{4}\zeta_{8}\theta\right) & \alpha\cdot\left(s_{5}+s_{6}\zeta_{8}+s_{7}\theta+s_{8}\zeta_{8}\theta\right)\\

  \zeta_{8}\bar{\alpha}\cdot\left(s_{5}+s_{6}\zeta_{8}+s_{7}\bar{\theta}+s_{8}\zeta_{8}\bar{\theta}\right)
  &
  \bar{\alpha}\cdot\left(s_{1}+s_{2}\zeta_{8}+s_{3}\bar{\theta}+s_{4}\zeta_{8}\bar{\theta}\right)\end{bmatrix}\]
and $\tau$ changes $\zeta_{8}$ into $-\zeta_{8}$.

\subsection{Four Relays, Single Antenna \label{one-antenna-n-relays}}
The generalization to $N=4$ relays is straightforward.  Codewords are
block diagonal matrices with $4$ blocks. Each block is a $2\times2$
matrix. Let $\mathbb{P}=\mathbb{Q}(i)$ and
$\mathbb{F}=\mathbb{Q}\left(\zeta_{16}\right)$ with
$\zeta_{16}=e^{\frac{i\pi}{8}}$ be an extension of $\mathbb{Q}(i)$ of
degree $4$. We choose
$\mathbb{K}=\mathbb{F}\left(\sqrt{5}\right)=\mathbb{Q}\left(\zeta_{16},\sqrt{5}\right)$.
We choose here, in order to preserve the shaping of the code,
$\gamma=\zeta_{16}$.  We prove in appendix \ref{zeta16-not-norm} that
$\zeta_{16}\notin N_{\mathbb{K}/\mathbb{F}}(\mathbb{K})$ and thus that
this code satisfies to the full rank and the NVD conditions.

\subsection{Single Relay, Two Antennas}

Since $n_s=2$ and $N=1$, we need a code whose codewords are
represented by a $4\times4$ NVD space-time code.  The $4\times4$
Perfect code of \cite{Oggier-1} satisfies to all criteria.

\subsection{Two Relays, Two Antennas}

We assume here that the source uses $2$ antennas and that there are
$2$ relays. The idea is to construct a $4\times4$ Perfect code not on
the base field $\mathbb{F}=\mathbb{Q}(i)$ as it is the case in
\cite{Oggier-1}, but on the base field $\mathbb{F}=\mathbb{Q}\left
  (\zeta_{8}\right )$. Thus, a rate-$2$ NVD code can be constructed as
follows:
\begin{itemize}
\item Take $\mathbb{F}=\mathbb{Q}\left (\zeta_{8}\right )$.
\item Choose $\mathbb{K}=\mathbb{F} \left( 2 \cos \left( \frac{2
        \pi}{15}\right) \right) =\mathbb{Q}\left(\zeta_{8},2 \cos
    \left( \frac{2 \pi}{15}\right)\right)$.
\item Finally, take $\gamma=\zeta_{8}$.
\end{itemize}

We can show, in the same way as in appendix \ref{zeta8-not-norm}, that
if $\zeta_8$ was a norm in $\mathbb{Q}\left(\zeta_{8},2 \cos \left(
    \frac{2 \pi}{15}\right)\right)$, then $i$ must be a norm in
$\mathbb{Q}\left(i,2 \cos \left( \frac{2 \pi}{15}\right)\right)$ which
contradicts the results of \cite{Oggier-1}. The case of $\gamma^2$ is
obvious since $\gamma^2=i$. Now, in order to prove that
$\gamma^3=i\zeta_{8}$ is not a norm, it is enough to replace
$\zeta_{8}$ by $i\zeta_{8}$ and $z$ by $y$ in appendix
\ref{zeta8-not-norm} and show in the same way that if $i\zeta_{8}$ was
a norm in $\mathbb{Q}\left(\zeta_{8},2 \cos \left( \frac{2
      \pi}{15}\right)\right)$, then $i$ must be a norm in
$\mathbb{Q}\left(i,2 \cos \left( \frac{2 \pi}{15}\right)\right)$.

\section{Numerical Results}
\label{sec:results}
In this section, we provide the simulation results on the performance
of some of the codes proposed in section~\ref{sec:some-examples}. The
performance is measured by the frame error rate~(FER) \vs receive SNR
per bit. For simplicity, we set the power allocation factors
$\pi_1=2\pi_2=2\pi_3$ for all the scenarios that we considered in this
section. An optimization on the $\pi_i$'s in function of $\rho$ and
$\SNR$ can improve the performance\footnote{A trivial suboptimal
  solution is to ``turn on'' the relay only when the $\rho$ and $\SNR$
  are high enough to give a better performance over the
  non-cooperative case.}. However, this kind of optimization is out of
the scope of this paper and will not be considered here. The
transmitted signal constellation is $4$- and $64$-QAM. The geometric
gain $\rho$ varies from $0$ to $20$\,dB.

\subsection{Single-Antenna Channel}
\label{sec:single-antenna-chann}
\begin{figure*}
  {\centerline{ \subfigure[$4$-QAM]
    {\includegraphics[width=13.5cm]{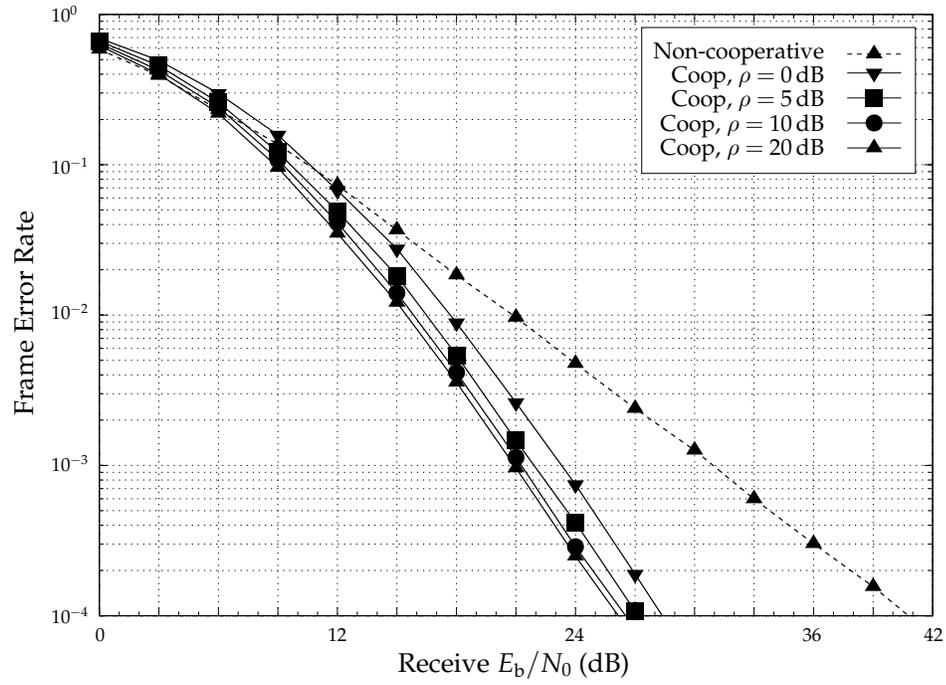}
\label{fig:r1_4QAM}}}
\centerline{\subfigure[$64$-QAM]
  {\includegraphics[width=13.5cm]{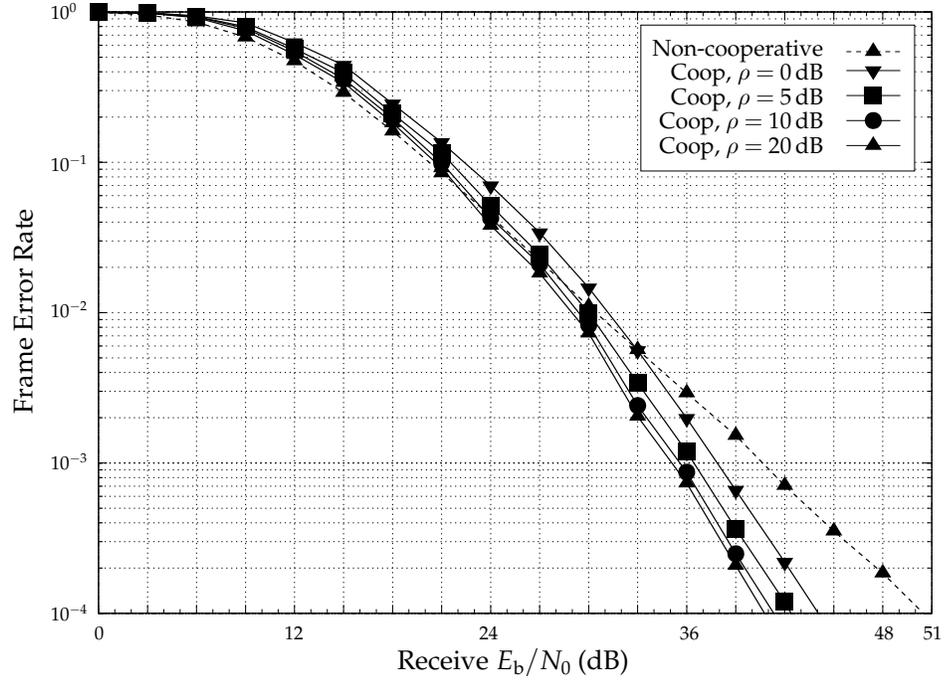}
\label{fig:r1_64QAM}}}}
\caption{Single-relay single-antenna  NAF channel, Rayleigh fading, Golden code.}
\label{fig:r1}    
\end{figure*}

Fig.~\ref{fig:r1} shows the performance of the Golden code on the
single-relay single-antenna channel. The performance of the channel
without relay is also shown in the figures.  In this case, the frame
length is $4$ symbols. Compared to the non-cooperative case, the
Golden code achieves diversity $2$. For $4$-QAM, a gain of
$12.5$\,dB~(\Resp $13.8,14.3$ and $14.8$\,dB) is observed for
$\rho=0$\,dB~(\Resp $5,10$ and $20$\,dB) at $\FER=10^{-4}$. First of
all, note that in the low SNR regime, the non-cooperative channel is
better than the cooperative channel. This is due to the error
cumulation~(at the relay) which is more significant than the diversity
gain provided by the relay in this regime. Then, we see that the
difference between $\rho=10$\,dB and $\rho=20$\,dB is negligible,
which means that a geometric gain of $10$\,dB is enough to achieve the
(almost)~best performance of the Golden code. In practice, it is often
possible to find this kind of ``helping agent'' (with a geometric gain
of $10$\,dB). When we increase the spectral efficiency ($64$-QAM),
same phenomena can be observed except that the gain of the relay
channel is reduced. Still, a gain of $6.3$ and $9.5$\,dB can be
obtained at $\FER=10^{-4}$ for $\rho=0$ and $20$\,dB.

\begin{figure*}
  {\centerline{ \subfigure[$4$-QAM]
    {\includegraphics[width=13.5cm]{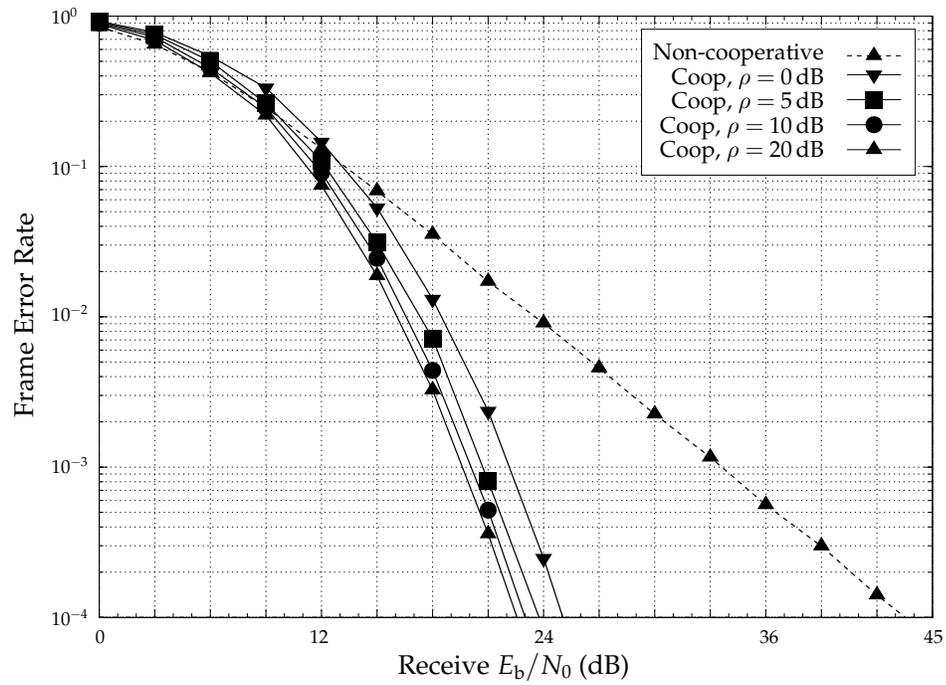}
\label{fig:r4_4QAM}}}
\centerline{\subfigure[$64$-QAM]
  {\includegraphics[width=13.5cm]{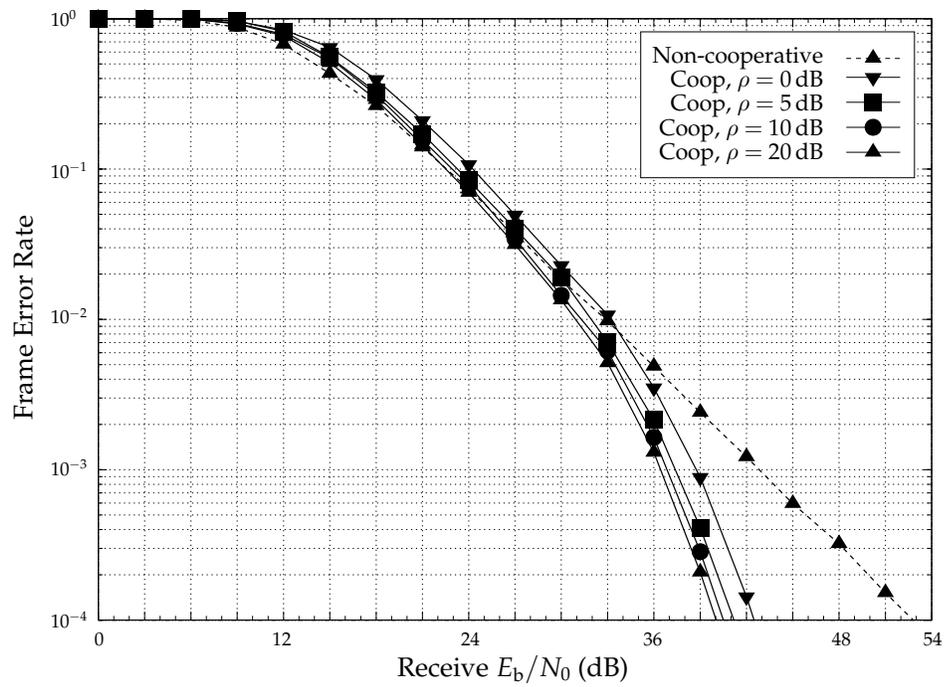}
\label{fig:r4_64QAM}}}}
\caption{Four-relay single-antenna NAF channel, Rayleigh fading, $\Ccal_{4,1}$.}
\label{fig:r4}    
\end{figure*}

The performance of $\Ccal_{4,1}$ on the four-relay single-antenna
channel is illustrated in Fig.~\ref{fig:r4}. The frame length is $16$.
For $\rho=20$\,dB, a gain of $21$\,dB~(\Resp $12.8$\,dB) at
$\FER=10^{-4}$ is obtained with $4$-QAM~(\Resp $64$-QAM).

\subsection{Multi-Antenna Channel}
\subsubsection{A $(1,2,1)$-Relay Channel}  
\begin{figure*}[!htbp]
  \begin{center}
    \includegraphics[angle=0,width=13.5cm]{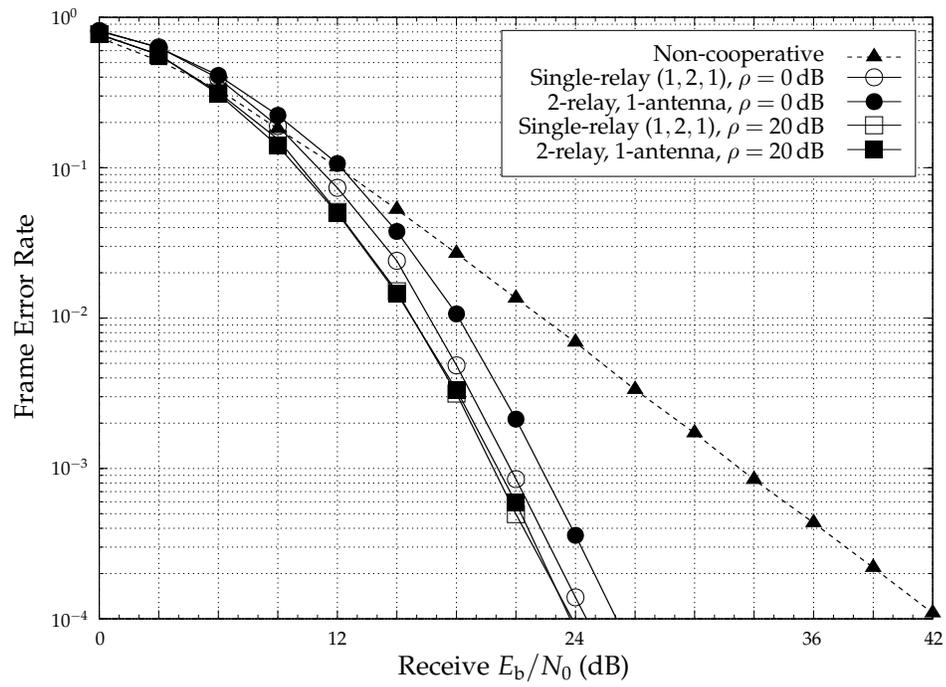}
    \caption{Single-relay $(1,2,1)$ NAF channel \vs two-relay single-antenna NAF channel, Rayleigh fading, 
      $\Ccal_{2,1}$ with $4$-QAM.}
    \label{fig:comp_121_211}    
  \end{center}
\end{figure*}

As an example for the case $n_r>n_s$, we consider the $(1,2,1)$
channel. Here, we use the virtual $N$-relay scheme with $N=2$. As
discussed in section~\ref{sec:n_r-geq-n_s}, the diversity order of
this scheme is comparable to the $2$-relay single-antenna channel. To
compare these two channels, we use the same code $\Ccal_{2,1}$. The
performance is shown in Fig.~\ref{fig:comp_121_211}. As compared to
the $2$-relay single-antenna channel, the $(1,2,1)$ channel has a gain
of $1.5$\,dB at $\FER=10^{-4}$ with $\rho=0$\,dB. With $\rho=20$\,dB,
the two channels have essentially the same performance.  In fact, the
inter-relay cooperation in the virtual two-relay channel improves the
receive SNR~($3$\,dB) of the source-relay channel with antenna
combining. Thus, the geometric gain $\rho$ is increased effectively.
However, as stated before, the global performace is not sensitive to
$\rho$ for large $\rho$'s. This is why there is a gain only with small
$\rho$'s.

\subsubsection{A $(2,2,2)$-Relay Channel}  
In the case of $n_s\geq n_r$, we consider a single-relay channel with
two antennas at each terminal. The code $\Ccal_{1,2}$ is actually the
$4\times4$ Perfect code. For the non-cooperative scenario, we take the
same code for fairness of comparison. More precisely, the
non-cooperative channel we consider here is equivalent to a
cooperative channel with $\pi_3=0$ and $\pi_1=\pi_2$. As shown in
\Fig{fig:r1_MIMO}, the gain of the cooperative channel over the
non-cooperative channel is much less significant in the SNRs of
interest. This is because the diversity order of the Perfect code in
the two-antenna non-cooperative channel is already $4$ and a diversity
gain does not play an important role in the scope of interest. Note
that at $\FER=10^{-5}$, the gain of the cooperative channel with
$\rho=0$\,dB over the non-cooperative channel is $2$\,dB for $4$-QAM
and $3$\,dB for $64$-QAM. Also note that the difference between
different $\rho$'s is within $1$\,dB.

\subsubsection{A $(2,2,2)$-Relay Channel with Shadowing}
\begin{figure*}
  {\centerline{ \subfigure[$4$-QAM]
    {\includegraphics[width=13.5cm]{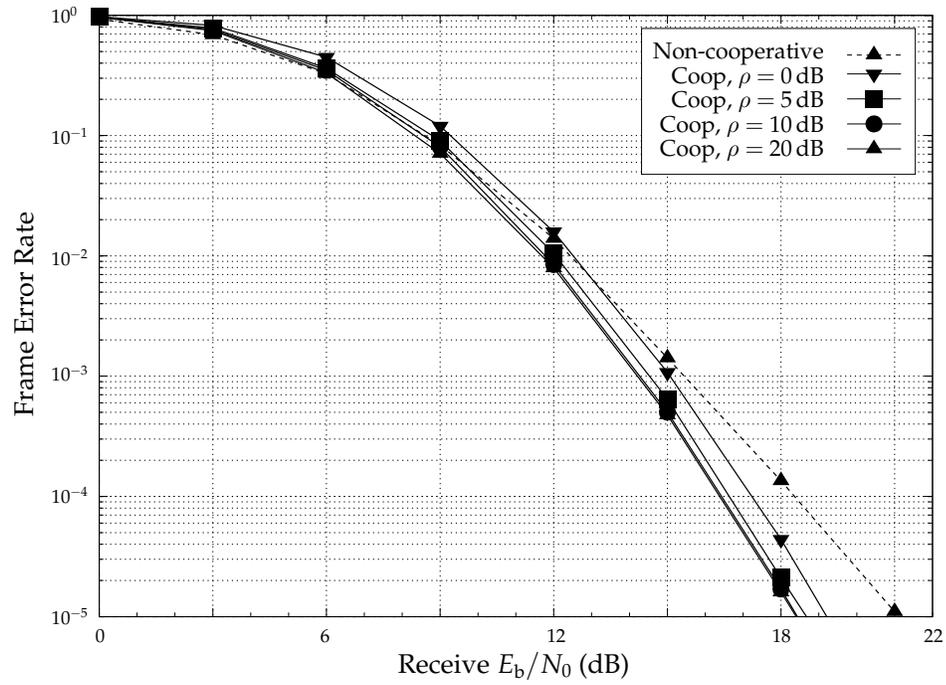}
\label{fig:r1_MIMO_4QAM}}}
\centerline{\subfigure[$64$-QAM]
  {\includegraphics[width=13.5cm]{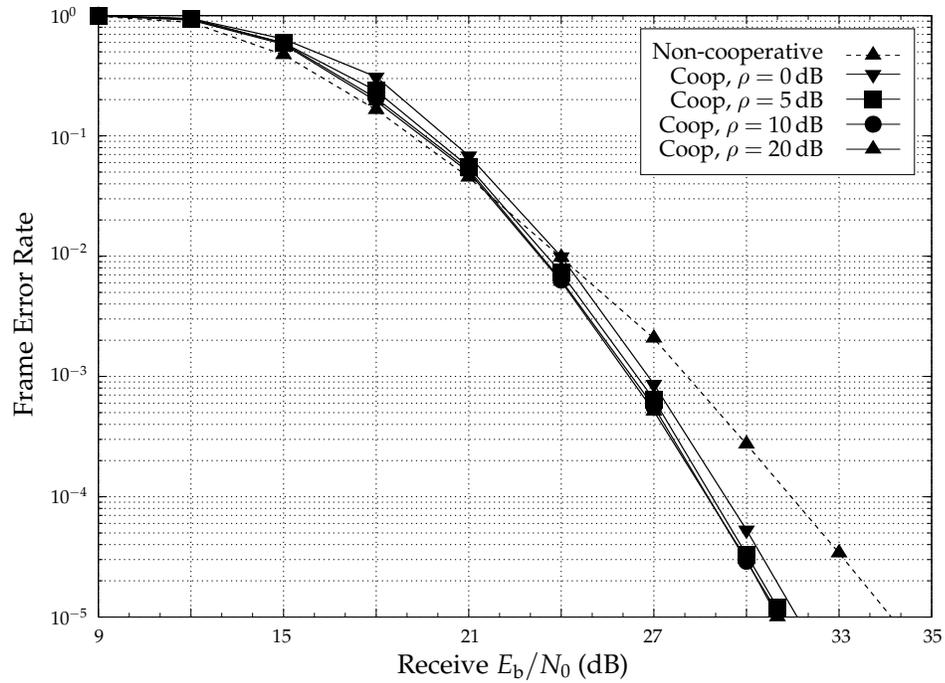}
\label{fig:r1_MIMO_64QAM}}}}
\caption{Single-relay $(2,2,2)$ NAF channel, Rayleigh fading, $4\times4$ Perfect code.}
\label{fig:r1_MIMO}    
\end{figure*}

\begin{figure*}[!htbp]
  \begin{center}
    \includegraphics[angle=0,width=13.5cm]{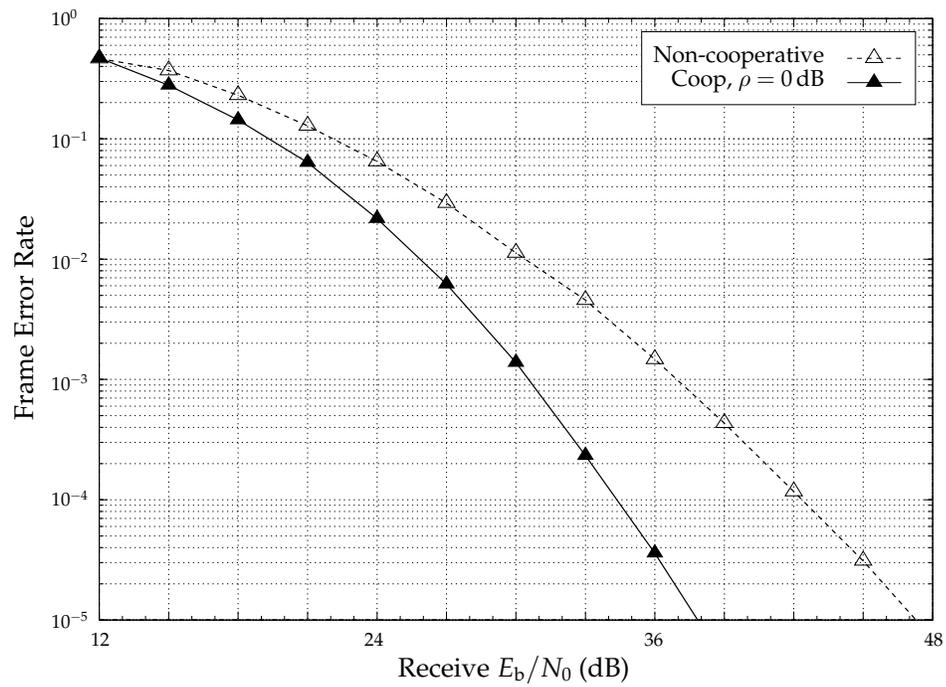}
    \caption{Single-relay $(2,2,2)$ NAF channel, Rayleigh fading, log-normal shadowing with variance $7$dB,  
      $4$-QAM, $4\times4$ Perfect code.  }
    \label{fig:MIMOshadowed}
  \end{center} \end{figure*}

In this scenario, we consider the shadowing effect of a wireless
channel. Assume that each link between terminals is shadowed.
Mathematically, the channel matrix is multiplied by a random scalar
variable, the shadowing coefficient. Suppose that this variable is
log-normal distributed of variance $7$\,dB\cite{Rappaport} and that
the shadowing is independent for different links.
Fig.~\ref{fig:MIMOshadowed} shows the performance of the cooperative
channel with the use of $\Ccal_{1,2}$~($4$-QAM) at $\rho=0$\,dB, as
compared to the non-cooperative channel. FER is the averaged frame
error rate on the channel fading and the shadowing. As shown in
\Fig{fig:MIMOshadowed}, the slope of the FER-SNR curve of the
non-cooperative channel is reduced as compared to the non-shadowing
case, in the scale of interest\footnote{With shadowing, the slope
  converges very slowly to the diversity order of the fading
  channel.}. Since the shadowing is independent between different
links, the cooperative channel mitigates the shadowing effect and we
get a larger gain over the non-cooperative channel than in the
non-shadowing case~(\Fig{fig:r1_MIMO_4QAM}). At $\FER=10^{-4}$, this
gain is $8$\,dB, in contrast to $1.2$\,dB in the non-shadowing
case~(\Fig{fig:r1_MIMO_4QAM}).

\section{Conclusion and Future Work}
\label{sec:conclusion}
In this paper, a half-duplex MIMO amplify-and-forward cooperative
diversity scheme is studied. We derived the optimal
diversity-multiplexing tradeoff of a MIMO Rayleigh product channel,
from which we obtain a lower bound on the optimal
diversity-multiplexing tradeoff of a MIMO NAF cooperative channel.
Moreover, we established a lower and upper bound on the maximum
diversity order of the proposed MIMO NAF channel and showed that they
coincide when the numbers of antenna satisfy certain conditions. Based
on the non-vanishing determinant criterion, we constructed a family of
short space-time block codes that achieve the DMT of our MIMO NAF
model. Our construction is systematic and applies to a system with
arbitrary number of relays and arbitrary number of antennas. Numerical
results on some explicit example codes revealed that significant gain
in terms of SNR can be obtained even with some non-optimized
parameters. This gain is much more important in the single-antenna
case than in the MIMO case. Fortunately, in reality, it is also the
case that we need cooperative diversity only when local antenna array
is not available.

Nevertheless, it still remains two important open problems to solve~:
\begin{enumerate}
\item optimization of the power allocation factors: Based on the
  statistical knowledge of the channel (notably $\rho$), how to choose
  the factors $\pi_i$'s in order to optimize the code performance
  according to certain criteria? We set $\pi_1=2\pi_2=2\pi_3$ for
  simplicity. It is clear that the optimal DMT is independent of these
  parameters. However, in practice, for different $\rho$ and $\SNR$,
  the factor $\pi_i$'s are significant for the performance~(\eg, the
  error rate performance).  How to analyze the impact theoretically is
  an interesting future work;
\item optimization of the matrix $\mB$: we set $\mB$ to be identity
  matrix and derived the lower bound~(\ref{eq:dNAFMIMO}). Based on the
  receiver CSI at the relay, is there an optimal matrix $\mB$ that
  gives a better DMT than the lower bound~(\ref{eq:dNAFMIMO})? This
  problem is independent of the code we use. Solving this problem may
  lead to solution for the exact DMT of the MIMO NAF channel.
\end{enumerate}

\appendices
\section{Preliminaries to the Proofs}

For sake of simplicity, we use the \emph{dot (in)equalites} throughout
the proofs to describe the behavior of different quantities in the
high SNR regime. More precisely,
\begin{itemize}
\item for probability related quantities,
  \begin{equation*}
    p_1 \asympteq p_2 \quad   \textrm{means} \quad \lim_{\SNR\to\infty}\frac{\log p_1}{\log \SNR} = \lim_{\SNR\to\infty}\frac{\log p_2}{\log \SNR};
  \end{equation*}%
\item for mutual information related quantities,
  \begin{equation*}
    \Ical_1 \asympteq \Ical_2 \quad \textrm{means} \quad \lim_{\SNR\to\infty}\frac{\Ical_1}{\log \SNR} = \lim_{\SNR\to\infty}\frac{\Ical_2}{\log \SNR};
  \end{equation*}%
\item for sets,
  \begin{equation*}
    \Scal_1 \asympteq \Scal_2 \quad \textrm{means} \quad \Prob{s\in\Scal_1}\asympteq\Prob{s\in\Scal_2}.
  \end{equation*}%
  $\asymptgeq$, $\asymptleq$, $\asymptsupseteq$ and $\asymptsubseteq$
  are similarly defined.
\end{itemize}

\begin{definition}[Exponential order\cite{ElGamal_coop}]
  For any nonnegative random variable $x$, the \emph{exponential
    order} is defined as
  \begin{eqnarray}
    \xi &\defeq& -\lim_{\SNR\to\infty} \frac{\log{x}}{\log\SNR}\label{def:expord}.
  \end{eqnarray}%
  We denote $x \asympteq \SNR^{-\xi}$.
\end{definition}

\begin{lemma}\label{lemma:PDFexp}
  Let $X$ be a $\chi^2$-distribution random variable with $2t$ degrees
  of freedom, the probability density function of its exponential
  order $\xi$ satisfies
  \yseqncasesasymptnnb{p_\xi}{\SNR^{-\infty}}{$\xi<0$;}{\SNR^{-\xi t}}{$\xi\geq0$.}%
  Let $\Scal$ be a certain set, $\xi_i$'s be independent random
  variables with $\xi_i\sim\chi^2_{2t_i}$, and
  $P_\Scal\defeq\prob\Big\{(\xi_1,\ldots,\xi_N)\in\Scal\Big\}$, then
  we have
  \begin{eqnarray*}
    P_\Scal \asympteq \SNR^{-d} \quad\textrm{with}\quad d=\inf_{(\xi_1\ldots{\xi_N})\in\Scal^+}\sum_{i=1}^N t_i \xi_i
  \end{eqnarray*}%
where $\Scal^+ = \Scal \bigcap \RR^{N+}$.
\end{lemma}

\section{Proof of Proposition~\ref{prop:prod}}\label{app:proof_prop1}
  \NC{\covH}{\mC_\mH}  
  In the high SNR regime, the outage probability is \cite{Zheng_Tse}
  \begin{equation} \label{eq:pout}
    \Pout(r\log\SNR) \asympteq \Prob{\log\det\left(1+\SNR\mA\transc{\mA}\right)<r\log\SNR}
  \end{equation}%
  with $\mA\defeq\mG\mH$. Let us define $\covH\defeq \mH\transc{\mH}$
  and $\mW\defeq \transc{\mA}\mA$. The entries of $\mG$ and $\mH$
  being \iid Rayleigh distributed, $\covH$ and $\mW|\mH$ are two
  central complex Wishart matrices~\cite{Tulino_Verdu}. Let
  $\mu_1>\cdots>\mu_l>0$ and $\lambda_1>\cdots>\lambda_l>0$ be the
  ordered eigenvalues of $\covH$ and $\mW$, then we have
  \cite{Ratnarajah,Tulino_Verdu}
\begin{equation*}
  \setlength{\nulldelimiterspace}{0pt}
  \left\{
      \begin{aligned}
        f(\mmu) &= G_{m,l}\prod_{k=1}^l \mu_k^{m-l} \prod_{k<p}^l (\mu_k-\mu_p)^2 \exp\left(-\sum_{k=1}^l \mu_k \right) \\
        f(\mlambda|\mmu) &= K_{l,n}\prod_{k=1}^l
        \mu_k^{l-n-1}\lambda_k^{n-l} \prod_{k<p}^l
        \frac{\lambda_k-\lambda_p}{\mu_k-\mu_p}
        \det\left[\exp\left(-\frac{\lambda_j}{\mu_i}\right)\right]
      \end{aligned}\right.
\end{equation*}%
with $K_{l,n}$ and $G_{m,l}$ being the normalization factors. Hence,
the joint pdf of $(\mlambda,\mmu)$ is
\begin{align*}
  f(\mlambda,\mmu) &= f(\mmu)f(\mlambda|\mmu) \\
  &= C_{l,m,n} \prod_{k=1}^l \mu_k^{m-n-1}\lambda_k^{n-l} \prod_{k<p}^l {(\lambda_k-\lambda_p)}{(\mu_k-\mu_p)} \\
  &\quad \cdot \exp\left(-\sum_{k=1}^l \mu_k\right)
  \det\left[\exp\left(-\frac{\lambda_j}{\mu_i}\right)\right],
\end{align*}%
where $C_{l,m,n}$ is the normalization factor. Define
$\alpha_i\defeq-\log\lambda_i/\log\SNR $ and
$\beta_i\defeq-\log\mu_i/\log\SNR$ for $i=1,\ldots,l$. Then, we have
\begin{align}
  f(\malpha,\mbeta) &= C_{l,m,n} (\log\SNR)^{2l}\prod_{k=1}^l \SNR^{-(n-l+1)\alpha_k} \SNR^{-(m-n)\beta_k} \nnb\\
  &\quad\cdot \prod_{k<p}^l {(\SNR^{-\alpha_k}-\SNR^{-\alpha_p})}{(\SNR^{-\beta_k}-\SNR^{-\beta_p})}  \nnb\\
  &\quad \cdot \exp\left(-\sum_{k=1}^l \SNR^{-\beta_k}\right)
  \det\left[\exp\left(-\SNR^{-(\alpha_j-\beta_i)}\right)\right].
  \label{eq:pdfjoint}
\end{align}%
First, we only consider $\beta_i\geq0,\forall i$, since otherwise,
$\exp\left(-\sum_{k} \SNR^{-\beta_k}\right)$ decays exponentially with
$\SNR$\cite{Zheng_Tse}. Then, we can show that
\begin{equation}
  \label{eq:detexp}
  \det\left[\exp\left(-\SNR^{-(\alpha_j-\beta_i)}\right)\right] \asympteq \SNR^{-\sum_{k=1}^l\sum_{i<k}(\alpha_i-\beta_k)^+}.
\end{equation}%
To see this, let us rewrite $D_l\defeq \det\left[\exp\left(-\SNR^{-(\alpha_j-\beta_i)}\right)\right]_{i,j=1}^l$ as
\begin{align*}
  D_l &= e^{-\sum_i\SNR^{-(\alpha_l-\beta_i)}} \det\matrix{
    e^{-\SNR^{-(\alpha_1-\beta_1)}+\SNR^{-(\alpha_l-\beta_1)}} &\cdots &e^{-\SNR^{-(\alpha_{l-1}-\beta_1)}+\SNR^{-(\alpha_l-\beta_1)}} & 1\\
    \vdots & \ddots& \vdots & \vdots \\
    e^{-\SNR^{-(\alpha_1-\beta_l)}+\SNR^{-(\alpha_l-\beta_l)}} &\cdots
    &e^{-\SNR^{-(\alpha_{l-1}-\beta_l)}+\SNR^{-(\alpha_l-\beta_l)}} &
    1
  }\\
     &\asympteq e^{-\SNR^{-(\alpha_l-\beta_l)}} \det \matrix {
     e^{-\SNR^{-(\alpha_1-\beta_1)}}-e^{-\SNR^{-(\alpha_1-\beta_l)}} &\cdots& e^{-\SNR^{-(\alpha_{l-1}-\beta_1)}}-e^{-\SNR^{-(\alpha_{l-1}-\beta_l)}} & 0 \\
     \vdots & \ddots& \vdots & \vdots \\
     e^{-\SNR^{-(\alpha_1-\beta_{l-1})}}-e^{-\SNR^{-(\alpha_1-\beta_l)}} &\cdots& e^{-\SNR^{-(\alpha_{l-1}-\beta_{l-1})}}-e^{-\SNR^{-(\alpha_{l-1}-\beta_l)}} & 0 \\
     e^{-\SNR^{-(\alpha_1-\beta_{l})}} & \cdots & e^{-\SNR^{-(\alpha_{l-1}-\beta_{l})}} & 1
     } \\
     &\asympteq e^{-\SNR^{-(\alpha_l-\beta_l)}} \det \matrix {
     e^{-\SNR^{-(\alpha_1-\beta_1)}}\left(1-e^{-\SNR^{-(\alpha_1-\beta_l)}}\right) &\cdots& e^{-\SNR^{-(\alpha_{l-1}-\beta_1)}}\left(1-e^{-\SNR^{-(\alpha_{l-1}-\beta_l)}}\right)\\
     \vdots & \ddots& \vdots\\
     e^{-\SNR^{-(\alpha_1-\beta_{l-1})}}\left(1-e^{-\SNR^{-(\alpha_1-\beta_l)}}\right) &\cdots& e^{-\SNR^{-(\alpha_{l-1}-\beta_{l-1})}}\left(1-e^{-\SNR^{-(\alpha_{l-1}-\beta_l)}}\right)
     } \\
     &=  e^{-\SNR^{-(\alpha_l-\beta_l)}} \prod_{i=1}^{l-1}\left(1-e^{-\SNR^{-(\alpha_i-\beta_l)}}\right)D_{l-1}
\end{align*}%
where the equations are obtained by iterating the identity
$\SNR^{-a}\pm\SNR^{-b} \asympteq \SNR^{-a}$ for $a<b$. Since
$1-e^{-x}\approx x$ for $x$ close to $0^+$, we have
$1-e^{-\SNR^{-(\alpha_i-\beta_l)}} \asympteq
\SNR^{-(\alpha_i-\beta_l)}$ if $\alpha_i>\beta_l$ and
$1-e^{-\SNR^{-(\alpha_i-\beta_l)}}\asympteq\SNR^0$ otherwise.  As
shown in the recursive relation above, we must have
$\alpha_i\geq\beta_i,\forall i$, so that $D_l$ does not decay
exponentially. In this case, we have
$e^{-\SNR^{-(\alpha_i-\beta_i)}}\asympteq\SNR^0,\forall i$. Thus, we
have $D_l \asympteq \SNR^{-\sum_{i<l}(\alpha_i-\beta_l)^+} D_{l-1}$,
and in a recursive manner, we get \Eq{eq:detexp}. Finally, we can
write the outage probability as
\begin{equation*}
  \Pout(r\log\SNR) = \int_{\Ocal(r)} f(\malpha,\mbeta) \d\malpha\d\mbeta
\end{equation*}%
where
\begin{equation*}
   \Ocal(r) \defeq \left\{(\malpha,\mbeta):\quad\sum_{k=1}^l(1-\alpha_k)^+ < r, {{\alpha_1\leq\cdots\leq\alpha_l,\atop\beta_1\leq\cdots\leq\beta_l},\alpha_i\geq\beta_i\geq0,\forall i}\right\}
\end{equation*}%
is the outage region in terms of $(\malpha,\mbeta)$ in the high SNR
regime. Since $\SNR^{-\alpha_k}-\SNR^{-\alpha_p}$(\Resp
$\SNR^{-\beta_k}-\SNR^{-\beta_p}$) is dominated by
$\SNR^{-\alpha_k}$(\Resp $\SNR^{-\beta_k}$) for $k<p$, from
\Eq{eq:detexp} and \Eq{eq:pdfjoint}, we have
\NC{\dab}{d_{\malpha,\mbeta}}
\begin{equation*}
  \Pout(r\log\SNR) \asympteq \int_{\Ocal(r)} \SNR^{-\dab}\d\malpha\d\mbeta
\end{equation*}%
with
\begin{equation*}
  \dab \defeq \sum_{i=1}^l (n-i+1)\alpha_i + \sum_{i=1}^l (m-n+l-i)\beta_i + \sum_{j=1}^{l}\sum_{i<j} (\alpha_i-\beta_j)^+. 
\end{equation*}%

Let $\Pout(r\log\SNR)\asympteq\SNR^{-\dA(r)}$. Then, we have 
\begin{equation}
  \dA(r) = \inf_{\Ocal(r)} \dab. \label{eq:optpbl}
\end{equation}%
The optimization problem \Eq{eq:optpbl} can be solved in two steps: 1)
find optimal $\mbeta$ by fixing $\malpha$, and then 2) optimize
$\malpha$.

Let us start from the feasible region
\begin{equation*}
  0 \leq \beta_1 \leq \alpha_1 \leq \beta_2 \leq \alpha_2 \leq \cdots \leq \beta_l\leq \alpha_l
\end{equation*}%
in which we have $\D\sum_{j=1}^{l}\D\sum_{i<j}(\alpha_i-\beta_j)^+ =
0$.  Note that the feasibility conditions require that $\beta_i$'s can
only move to their left in terms of their positions relative to the
$\alpha_i$'s and that $\beta_i$ can never be on the left of $\beta_j$
for $i>j$. Let $b_i$ denote the coefficients of $\beta_i$ in $\dab$.
The initial values of $b_i$'s are $b_i^{(0)}\defeq m-n+l-i$ where
$b_1^{(0)}>b_1^{(0)}>\cdots>b_l^{(0)}$. As long as $b_i$ is positive,
$\beta_i$ should decrease~(pass the $\alpha_j$ just left to it) to
make the objective function $\dab$ smaller, with the feasibility
conditions being respected. Each time a $\beta_i$ passes a $\alpha_j$
from right to left with $j<i$, $b_i$ decreases by $1$. When $b_i=0$,
$\beta_i$ should stop decreasing.  Therefore, the optimal region is
such that
\begin{equation*}
  c_i \defeq b_i^{(0)} - b_i^* = \left[\min\left\{i-1,b_i^{(0)}\right\}\right]^+,\quad i=1,\ldots,l.
\end{equation*}
For $i$ such that $b_i^{(0)}<0$, $\beta_i^*=\alpha_i$. For
$b_i^{(0)}\geq0$, $\beta_i^*$ is $\alpha_{i-1-c_i}$ if $i-1-c_i\geq 1$
and $0$ otherwise. Note that $b_i^*$'s are independent of $\alpha_i$'s
and only depend on $(m,n,l)$. This is why we can separate the
optimization problem into two steps. After replacing the optimal
$\mbeta$ in $\dab$ and some basic manipulations, we obtain
\begin{align}
  d_{\malpha} &= \sum_{k=1}^{l-\Delta} \left(q+1-2k+\left\lfloor\frac{l+k+\Delta}{2}\right\rfloor\right)\alpha_k + \sum_{k=l-\Delta+1}^{l} \left(q+l+1-2k\right) \alpha_k \nnb\\
  &= \trans{\ma} \malpha \label{eq:da}
\end{align}%
where $a_k$ is non-negative and is non-increasing with $k$. Hence, the
optimal solution is $\alpha_k=1,k=s+1,\ldots,l$ and
$\alpha_k=0,k=1,\ldots,s$, from which we have $\dA(s) = \sum_{s+1}^l
a_k$. For $s\geq l-\Delta$,
\begin{align}
  \dA(s) &= \sum_{k=s+1}^l q+l+1-2k \nnb\\
  &= (q-s)(l-s). \label{eq:x}
\end{align}%
For $s\leq l-\Delta-1$,
\begin{align}
  \dA(s) &= \sum_{k=s+1}^{l-\Delta} \left(q+1-2k+\left\lfloor\frac{l+k+\Delta}{2}\right\rfloor\right)\alpha_k + \sum_{k=l-\Delta+1}^{l} \left(q+l+1-2k\right) \alpha_k \nnb\\
  &= (q-s)(l-s) - \frac{1}{2}\left\lfloor
    \frac{(l-\Delta-s)^2}{2}\right\rfloor. \label{eq:xx}
\end{align}%
By combining \Eq{eq:x} and \Eq{eq:xx}, we get \Eq{eq:dmt_pr}.

\section{Proof of Theorem~\ref{thm:dmt_MIMONAF}}\label{app:proof_thm1}
The main idea of the proof is to get lower bounds on the DMT by
lower-bounding the mutual information of the channel defined by
\Eq{eq:channel_MIMO} and \Eq{eq:channelH_MIMO}. Since the
multiplicative constants have no effects on the DMT, for simplicity of
demonstration, we will neglect them and rewrite $\mSigma=\Id +
\mP\transc{\mP}$ and
\begin{eqnarray*}
  \He &=& \matrix{\mF & \mbs{0} \\\mSigmasqrt\mP\mH & \mSigmasqrt\mF}.
\end{eqnarray*}%
The mutual information of the channel $\He$ is
\begin{align}
  \Ical(\mx;\He\mx+\mz) &\asympteq
  \log\det\left(\mI+\SNR\He\transc{\He}\right). \nnb
\end{align}%

\begin{lemma}\label{lemma:Sigma}
  Let $\succeq$ be the generalized inequality for matrices\footnote{
    $\mA\succeq\mB$ means that $\mA-\mB$ is positive semidefinite.},
  then
  \begin{equation}
    \Id \succeq \inv{\mSigma} \succeq \inv{\left(1+\lmax(\mP\transc{\mP})\right)}\cdot\Id\label{eq:Sigma}
  \end{equation}%
  and there exists a matrix $\mB$ satisfying the power constraint
  \Eq{eq:constraintB} such that
  \begin{equation}
    1+\lmax(\mP\transc{\mP}) \asympteq \SNR^0 \label{eq:lmaxPP}.
  \end{equation}
\end{lemma}
\begin{proof}
  \Eq{eq:Sigma} comes from the definition of $\mSigma$. \Eq{eq:lmaxPP}
  can be shown by construction. Let us take
  \begin{equation}
    \sqrt{\SNR}\mB \defeq \sqrt{c\cdot\min{\left\{\lmax^{-1}\left(\mH\transc{\mH}\right), 1\right\}}}\cdot\Id.\label{eq:B}
  \end{equation}%
  Then, the power constraint \Eq{eq:constraintB} is always satisfied
  with $c = \inv{\left(\SNR^{-1}+\pi_1\rho\right)}$. Since
  $c\asympteq\SNR^0$, we have
\begin{equation}\label{eq:PP}
  1 \leq 1+\lmax{(\mP\transc{\mP})} \asymptleq 1+\SNR^{-\alpha_{\max}} \asympteq \SNR^0
\end{equation}%
where $\alpha_{\max}$ is the exponential order of
$\lmax(\transc{\mG}\mG)$ and is positive with probability $1$ in the
high SNR regime\cite{Zheng_Tse,ElGamal_coop}.
\end{proof}

By \Lemma{lemma:Sigma} and the concavity of $\log\det(\cdot)$ on
positive matrices, we have
\begin{equation*}
  \log\det\left(\Id+\SNR\mHh\transc{{\mHh}}\right) \geq 
  \log\det\left(\Id+\SNR{\mHt}\transc{{\mHt}}\right) \geq
  \log\det\left(\Id+\SNR(1+\lmax{(\mP\transc{\mP})})^{-1}{\mHh}\transc{{\mHh}}\right)
\end{equation*}%
with 
\begin{equation*}
  {\mHh} \defeq \matrix{\mF & \mZero \\\mP\mH & \mF}.
\end{equation*}
Therefore, with $\mB$ in \Eq{eq:B}, we have $
\log\det\left(\Id+\SNR{\mHt}\transc{{\mHt}}\right) \asympteq
\log\det\left(\Id+\SNR{\mHh}\transc{{\mHh}}\right)$.  Assume that in
the rest of the proof, we always consider $\mB$ being in the form
\Eq{eq:B}. Then we have \newcommand{\Imax}{\Ical_{\max}}
\begin{equation}
  \Imax\defeq \max_{\mB\in\Bcal} \Ixy \asymptgeq \log\det\left(\Id+\SNR{\mHh}\transc{{\mHh}}\right)\label{eq:Ixy}
\end{equation}%
where $\Bcal$ is the set of matrices $\mB$ that satisfy the power
constraint \Eq{eq:constraintB}. Define $\mM\defeq
\Id+\SNR{\mHh}\transc{{\mHh}}$, we have
\begin{equation*}
  \mM \defeq \matrix{\Id+\SNR\mF\transc{\mF} & \SNR\mF\transc{\mH}\transc{\mP} \\ \SNR\mP\mH\transc{\mF} & \Id+\SNR\left(\mF\transc{\mF}+\mP\mH\transc{\mH}\transc{\mP}\right)}.
\end{equation*}%
Using the identity 
\begin{equation*}
  \det\left(\matrix{\mA &\mB \\\mC&\mD}\right) = \det(\mA)\det(\mD-\mC\inv{\mA}\mB)
\end{equation*}%
and some basic manipulations, we have
\begin{align}
  \det(\mM) 
  &= \det(\Id+\SNR\mF\transc{\mF}) \det\left( \Id + \SNR\mF\transc{\mF} +
    \SNR\mP\mH\mOmega\transc{\mH}\transc{\mP}\right)\label{eq:detM}
\end{align}%
where $\mOmega \defeq \Id -
\SNR\transc{\mF}\inv{(\Id+\SNR\mF\transc{\mF})}\mF$ is positive
definite. By the matrix inversion lemma $\inv{(\Id+\mL\mC\mR)} = \Id -
\mL\inv{(\mR\mL+\inv{\mC})}\mR$, we have
\begin{equation*}
  \mOmega = \inv{(\Id+\SNR\transc{\mF}\mF)}.  
\end{equation*}%
From \Eq{eq:Ixy} and \Eq{eq:detM}, we can obtain two lower bounds on
$\Imax$. The first one is
\begin{align}
  \Imax &\asymptgeq 2\log\det(\Id+\SNR\mF\transc{\mF}),
       \label{eq:lb11}
\end{align}%
whereas the second one is
\begin{align}
  \Imax &\asymptgeq \log\det(\Id+\SNR\mF\transc{\mF}) + \log\det(\Id+\SNR\mP\mH\mOmega\transc{\mH}\transc{\mP}) \nnb\\
  &=    \log\det(\Id+\SNR\transc{\mF}\mF) + \log\det(\Id+\SNR\mOmega\transc{\mH}\transc{\mP}\mP\mH) \nnb\\
  &=    \log\det(\Id+\SNR\transc{\mF}\mF + \SNR\transc{\mH}\transc{\mP}\mP\mH) \nnb\\
  &\geq \log\det(\Id + \SNR\transc{\mH}\transc{\mP}\mP\mH)
  \label{eq:tmp}.
\end{align}%
Since in \Eq{eq:B},
$\min{\left\{\lmax^{-1}\left(\mH\transc{\mH}\right), 1\right\}}
\asympteq \SNR^0$, from \Eq{eq:def_P} and \Eq{eq:tmp}, we have
\begin{equation}
  \label{eq:lb2}
  \Imax \asymptgeq \log\det\left(\Id + \SNR\transc{\mH}\transc{\mG}\mG\mH\right).
\end{equation}

The outage probability is
\begin{IEEEeqnarray*}{rCl}
  \Prob{\Imax<2r\log\SNR} &\asymptleq& \textrm{Prob}\Biggl\{{2\log\det(\Id+\SNR\mF\transc{\mF})\leq2r\log\SNR,\atop \log\det(\Id + \SNR\transc{\mH}\transc{\mG}\mG\mH) \leq 2r\log\SNR}\Biggr\}\nnb\\
  &=& \textrm{Prob}\Bigl\{2\log\det(\Id+\SNR\mF\transc{\mF})\leq2r\log\SNR\Bigr\}\\
  & & \cdot\textrm{Prob}\Bigl\{\log\det(\Id + \SNR\transc{\mH}\transc{\mG}\mG\mH) \leq 2r\log\SNR\Bigr\} \nnb\\
  &\asympteq& \SNR^{-\bigl(d_{\mF}(r)+d_{\mG\!\mH}(2r)\bigr)}
\end{IEEEeqnarray*}%
where the second line follows from the independency between $\mF$
and $\mG\mH$.

\section{Proof of Theorem~\ref{thm:dmt_MIMONAF_N} and Corollary~\ref{coro:1}}\label{app:proof_thm2}

\subsection{Proof of Theorem~\ref{thm:dmt_MIMONAF_N}}
As in the case of the single-relay channel, we need two lower bounds
on the mutual information. Since the mutual information of the
$N$-relay channel is the sum of that of the $N$ single-relay channels,
these two lower bound can be obtained directly from \Eq{eq:lb11} and
\Eq{eq:lb2}
\begin{equation*}
  \setlength{\nulldelimiterspace}{0pt}
      \begin{IEEEeqnarraybox}[\relax][c]{l'l'l}
        \Ixy &\asymptgeq& 2N\log\det(\Id+\SNR\mF\transc{\mF}) \\
        \Ixy &\asymptgeq& \sum_{i=1}^N \log\det(\Id + \SNR\transc{\mH}_i\transc{\mG}_i\mG_i\mH_i)
      \end{IEEEeqnarraybox}
\end{equation*}%
The outage probability is upper bounded by
  \begin{multline}
    \Prob{\Ixy<2Nr\log\SNR} \\ \leq
    \textrm{Prob}\Biggl\{{2N\log\det(\Id+\SNR\mF\transc{\mF})\leq2Nr\log\SNR,\atop\sum_{i}
      \log\det(\Id + \SNR\mHtrc[i]\mGtrc[i]\mG_i\mH_i) \leq
      2Nr\log\SNR}\Biggr\}.
  \label{eq:secondprob}    
  \end{multline}
  
  \newcommand{\ai}{\malpha_{i}} \newcommand{\dai}{d_{\ai}}
  \newcommand{ \aiset}{\left\{\ai\right\}_{i=1}^N}
  \newcommand{\drelay}{d_{\textrm{relay}}}
  \newcommand{\Or}{\Ocal_g(r)} \newcommand{\Br}{\Bcal_g(r)}
  \newcommand{\Oir}[1][r]{\Ocal_i(#1)}
  \newcommand{\Bir}[1][r]{\Bcal_i(#1)}
  
  Let us denote $\Ical_i\defeq \log\det(\Id +
  \SNR\mHtrc[i]\mGtrc[i]\mG_i\mH_i)$ and $\ai$ the set of exponential
  orders of the ordered eigenvalues of
  $\transc{\mH}_i\transc{\mG}_i\mG_i\mH_i$. The pdf of $\ai$ is
  $p_{\ai} \asympteq \SNR^{-\dai}$ where from \Eq{eq:da}, $\dai$ is
  nondecreasing with respect to the component-wise inequality, \ie,
\begin{equation}
  \label{eq:nondec}
  d_{\ai'}\geq d_{\ai}\ \textrm{if}\ \ai'\succeq \ai.  
\end{equation}%
Let us define
\begin{eqnarray*}
  \Or &\defeq& \left\{ \aiset~:~
  \sum_{i=1}^N\sum_{k=1}^q\left(1-\alpha_{i,k}\right)^+\leq 2Nr \right\}, \\
  \Oir &\defeq& \left\{ \ai~:~
  \sum_{k=1}^q\left(1-\alpha_{i,k}\right)^+\leq r \right\}. 
\end{eqnarray*}%
Then, the outage probability is
\begin{align*}
  \Prob{\sum_{i=1}^N \Ical_i \leq2Nr\log\SNR} &\asympteq \Prob{\aiset\in\Or} \\
  &\asympteq \SNR^{-\drelay}
\end{align*}%
where $\drelay$ is
\begin{align*}
  \drelay &= \inf_{\Or} \sum_{i=1}^N \dai \\
  &= \inf_{\mtheta:\ \sum_i \theta_i=1} \left(\sum_{i=1}^N \inf_{\Oir[2N\theta_i r]} \dai \right) \\
  &= \inf_{\mtheta:\ \sum_i \theta_i=1} \left(\sum_{i=1}^N
    d_{\mG_i\mH_i}(2N\theta_i r) \right)
\end{align*}
with the second equality from the fact that the minimal elements lie
always in the boundary when \Eq{eq:nondec} is true. For $\mG_i\mH_i$
identically distributed for all $i$, \Eq{eq:lb3} is obtained by the
convexity of $d_{\mG\mH}$.

\subsection{Proof of Corollary~\ref{coro:1}}
For simplicity, we prove the particular case $N=1$ here. For $N>1$,
same method applies. The lower bound is a direct consequence of
\Thm{thm:dmt_MIMONAF_N}. The upper bound can be found by relaxing the
half duplex constraint, \ie, $\He \defeq \mSigmasqrt \left[{\mP\mH \ 
    \mF}\right]$ with all matrices being similarly defined as before.
Define ${\mHh}\defeq \left[ \mP\mH \ \mF \right]$. First, since
$\Id\succeq\mSigmasqrttrc\,\mSigmasqrt$, we have
  \begin{align}
    \Ical(\mx;\sqrt{\SNR}\He\mx+\mz) &= \log\det\left(\Id+\SNR\transc{\mHh}\mSigmasqrttrc\,\mSigmasqrt\mHh\right) \nnb\\
    &\leq \log\det\left(\Id+\SNR\mF\transc{\mF}+\SNR\mP\mH\mHtrc\mPtrc\right) \label{eq:5} \\
    &=\log\det\left(\Id+\SNR\mF\transc{\mF}+\SNR\mG(\SNR\mB\mH\mHtrc\mBtrc)\mGtrc\right) \label{eq:7} \\
    &\asymptleq\log\det\left(\Id+\SNR\mF\transc{\mF}+\SNR\mG\mGtrc\right) \label{eq:7} 
  \end{align}%
  which means that the channel $\He$ is asymptotically worse than the
  channel $\left[{\mG \quad \mF}\right]$ in the high SNR regime. Thus,
  we have $d_{\mF}(0) + d_{\mG}(0) \geq d_{\NAF}(0)$.
  
  Then, since $\Id\succeq\mSigmasqrttrc\mPtrc\mP\mSigmasqrt$, another
  bound is
  \begin{align}
    \Ical(\mx;\sqrt{\SNR}\He\mx+\mz) &= \log\det\left(\Id+\SNR\mSigmasqrt\mF\transc{\mF}\,\mSigmasqrttrc + \SNR\mSigmasqrt\mP\mH\mHtrc\mPtrc\,\mSigmasqrttrc\right) \label{eq:8} \\
    &\leq \log\det\left(\Id+\SNR\mSigmasqrt\mF\transc{\mF}\,\mSigmasqrttrc\right) + \log\det\left(\Id+\SNR\mSigmasqrt\mP\mH\mHtrc\mPtrc\,\mSigmasqrttrc\right) \label{eq:8} \\
    &\leq \log\det\left(\Id+\SNR\transc{\mF}\mF\right)+\log\det\left(\Id+\SNR\mHtrc\mH\right) \label{eq:9} 
  \end{align}%
  from which we have $d_{\mF}(0) + d_{\mH}(0) \geq d_{\NAF}(0)$.
  
  When the channel is Rayleigh, \Prop{prop:prod} applies. As indicated
  in remark~1, we have $d_{\mG\mH} =
  \min\bigl\{d_{\mG},d_{\mH}\bigr\}$ for $\Abs{m-n}\geq l-1$ and the
  lower bound and the upper bound in \Eq{eq:max_d} match.

\NC{\Hgram}{\mH\transc{\mH}}
\NC{\li}{\lambda_i}
\RNC{\ai}{\alpha_i}
\section{Proof of Theorem~\ref{thm:dmt}}\label{app:proof_thm3}

To prove \Thm{thm:dmt}, it is enough to show that in the high SNR
regime, an error occurs with the rate-$n$ NVD code $\Xcal$ only when
the channel is in outage for a rate $\frac{q}{n}r$. To this end, we
will show that the error event set of $\Xcal$ is actually included in
the outage event set, in the high SNR regime.

\subsection{Outage event}
\label{sec:outage-event}

For a channel $\mH$, the outage event at high SNR is~\cite{Zheng_Tse}
\begin{equation*}
  \Oe(r) \asympteq \big\{\mH~:~\log\det\left(\Id+\SNR\mH\transc{\mH}\right) < r \log\SNR\big\}.
\end{equation*}%
Let us develop the determinant as\footnote{To see this, consider the
  identity $\D \det(\mM-x\Id) = (-1)^n \prod_{i=1}^n(x-\lambda_i)$
  where $\lambda_i$'s is the eigenvalues of $\mM$.}
\begin{equation*}
  \det(\Id+\SNR\Hgram) = 1+\sum_{i=1}^{q}\SNR^i D_i(\Hgram) 
\end{equation*}%
where $D_i(\mM)$ is the sum of $\binom{q}{i}$ products of $i$
different eigenvalues of $\mM$. In particular, we have $D_1(\mM) =
\Tr(\mM)$ and $D_n(\mM)=\det(\mM)$. Let $\lambda_i$ denote the $i^\th$
smallest eigenvalue of $\Hgram$ and $\alpha_i$ denote the exponential
order of $\lambda_i$, \ie, $\li\asympteq\SNR^{-\ai}$ with
$\alpha_1\geq\alpha_2\geq\cdots\geq\alpha_{q}$. Then, we have
\begin{equation*}
  D_i \asympteq \SNR^{-\sum_{k=q-i+1}^q\alpha_k} \quad \for\ i = 1,\ldots, q
\end{equation*}%
since $\sum_{k=q-i+1}^q \alpha_k$ is the smallest among all the
combinations of $i$ different $\alpha$'s. Now, we are ready to write
\begin{align}
  \Oe(r) &\asympteq \left\{\mH~:~1+\sum_{i=1}^{q} \SNR^i D_i(\Hgram) < \SNR^r \right\} \nnb\\
  &\asympteq \Bigl\{\mH~:~\SNR^i D_i(\Hgram) \asymptleq \SNR^r, \quad \forall i=1,\ldots, q \Bigr\} \nnb\\
  &\asympteq \left\{\malpha~:~i-\left(\sum_{k=q-i+1}^{q} \alpha_k\right) \leq r, \quad \forall i=1,\ldots, q \right\} \nnb\\
  &= \left\{\malpha~:~\sum_{k=j+1}^{q} \alpha_k \geq  (q-j)-r, \quad \forall j=0,\ldots, q-1 \right\}. \nnb\\
\end{align}

\subsection{Error event of a rate-$n$ NVD code}
\label{sec:error-event-rate}

Let us now consider the error event of a rate-$n$ NVD code $\Xcal$.
We will follow the footsteps of \cite{Elia}. Using the sphere bound,
the error event of ML decoding conditioned on a channel realization
$\mH$ is
\begin{align}
  \Ecal_{\mH} &\subseteq \left\{\mW~:~\Frob{\mW}>\frac{\dEmin}{4}\right\} \nnb\\
  &\asympteq \Bigl\{w~:~-w\geq\eta \Bigr\} \nnb
\end{align}%
where $\mW$ is the AWGN matrix with \iid entries; $d_{\min}$ is the
minimum Euclidean distance between two received codewords, \ie,
$\dEmin\defeq\min{\Frob{\mH\Delta_\mX}}$; $w$ is the exponential order
of $\Frob{\mW}$~($\sim\chi^2_{2n_R n_T}$) and $\eta$ is that of
$1/\dEmin$. Therefore, the error probability conditioned on $\mH$ is
\begin{equation*}
  \PEH \asymptleq \textrm{Prob}\bigl\{-w\geq\eta\bigr\} \asympteq \SNR^{-\dEH}
\end{equation*}%
where by \Lemma{lemma:PDFexp}, we have
\yseqncases{\dEH}{\inf_{w\in\RR^+}\nR \nT w = 0}{$\eta\leq0$;}{\infty}{$\eta>0$.}%
Then the average error probability becomes
\begin{align}
  \PE &= \int \PEH\PHH\d\mH \nnb\\
  &\asymptleq \int_{\eta\leq0} \PHH \d\mH\nnb\\
  & = \textrm{Prob}\bigl\{\eta\leq0\bigr\}. \nnb
\end{align}%
Therefore, we get the error event in the high SNR regime
\begin{equation}
\label{eq:allbound}
  \Ecal \asymptsubseteq \bigl\{\malpha~:~\eta\leq0\bigr\} \subseteq \bigcap_{\xi\leq\eta}\bigl\{\malpha~:~\xi\leq0\bigr\}
\end{equation}%
with $\xi$ being any lower bound on $\eta$. Using the same arguments
as in \cite{Elia}, with a \nntcode, we can get $q$ lower bounds on
$\dEmin$
\begin{equation*}
  \SNR^{\eta} \asympteq \dEmin(\malpha) \asymptgeq \SNR^{\delta_j(\malpha)}, \quad j=0,\ldots,q-1
\end{equation*}%
with
\begin{equation}
  \label{eq:lb_dmin}
  \delta_j(\malpha) = 1- \frac{q}{n} \frac{r}{j+1} - \sum_{i=q-j}^{q} \frac{\alpha_i}{j+1}.
\end{equation}%
Finally, from \Eq{eq:allbound} and \Eq{eq:lb_dmin}, we get
\begin{align}
  \Ecal &\asymptsubseteq \Bigl\{\malpha~:~\delta_j\leq0, \quad \forall j=0,\ldots, q-1 \Bigr\} \nnb\\
  &= \left\{\sum_{k=j+1}^{q} \alpha_k \geq  (q-j)-\frac{q}{n}r, \quad \forall j=0,\ldots, q-1 \right\} \nnb\\
  &\asympteq \Oe\left({\frac{q}{n}r}\right)\nnb
\end{align}
which implies that 
\begin{equation*}
  \dX(r) \geq \dout\left(\frac{q}{n}r\right).
\end{equation*}

\NC{\PoutNAF}{P_{\out}^{\NAF,N}}
\NC{\PoutH}{P_{\out}^{\mH}}
\section{Proof of Theorem~\ref{thm:achievability}}\label{app:proof_thm4}
Consider the channel $\mLambda = \diag({\mHt}_1,\ldots,{\mHt}_N)$ with
${\mHt}_i$ being similarly defined as $\He$ in \Eq{eq:channelH_MIMO}
except that $\mG,\mH,\mB$ in \Eq{eq:channelH} are replaced by
$\mG_i,\mH_i,\mB_i$, respectively. Since one channel use of $\mLambda$
is equivalent to $2N$ channel uses of an $N$-relay NAF channel, \ie,
  \begin{equation}
    C_{\mLambda} = 2N\CNAF \label{eq:capa}
  \end{equation}%
  where $C_{\mLambda}$ and $\CNAF$ are the capacities of the channel
  $\mLambda$ and the equivalent $N$-relay NAF channel, measured by
  bits per channel use.  Therefore, we have
  \begin{equation}
    d_{\NAF,N}(r) = d_{\NAF,N}^{\out}(r) = d_{\mLambda}^{\out}(2Nr)\label{eq:dN}
  \end{equation}%
  where the first equality comes from the fact that the outage upper
  bound of the tradeoff can be achieved~\cite{ElGamal_coop} and the
  second comes from \Eq{eq:capa} and the definition of outage since
  \begin{equation*}
    \Pout^{\mLambda}(2Nr) \defeq \textrm{Prob}\bigl\{C_{\mLambda}<2Nr\log\SNR\bigr\} = \textrm{Prob}\bigl\{\CNAF< r\log\SNR\bigr\} = \PoutNAF(r).
  \end{equation*}
  
  On the other hand, by using a code $\Ccal$ defined above, an
  equivalent channel model of the $N$-relay channel is
  \begin{equation*}
    \mY = \sqrt{\SNR}\,\mLambda \mX + \mZ
  \end{equation*}%
  with $\mX\in \Xcal$. By \Thm{thm:dmt}, we have
\begin{equation}
  d_{\Ccal}(r) = d_{\Xcal}(2r) \geq d_{\mLambda}^{\out}(2Nr). \label{eq:dCcal}
\end{equation}%
From \Eq{eq:dN} and \Eq{eq:dCcal}, we obtain
\begin{equation*}
  d_{\Ccal}(r) \geq d_{\NAF,N}(r).
\end{equation*}

\section{$\zeta_{8}$ is not a norm in $\mathbb{Q}\left(\zeta_{8},\sqrt{5}\right)$\label{zeta8-not-norm}}
\begin{figure}[ht]
\begin{center}\includegraphics[%
  width=0.4\textwidth]{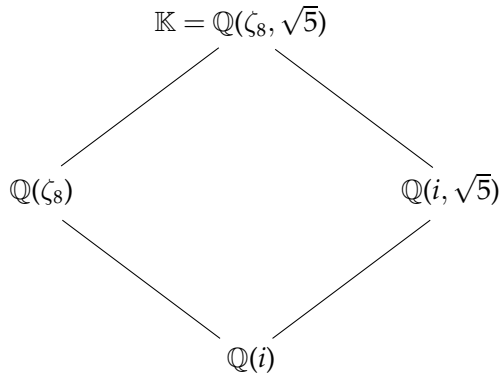}\end{center}
\caption{\label{cap:fields-extension}Two ways of extending $\mathbb{Q}(i)$
up to $\mathbb{K}$.}
\end{figure}

We prove, in this appendix, that $\zeta_{8}$ is not a norm of an
element of $\mathbb{K}=\mathbb{Q}\left(\zeta_{8},\sqrt{5}\right)$.
Assume that $\zeta_{8}$ is a norm in $\mathbb{K}$, \ie,
\begin{equation} \exists
  x\in\mathbb{K},N_{\mathbb{K}/\mathbb{Q}\left(\zeta_{8}\right)}(x)=\zeta_{8}.\label{eq:x_existence}\end{equation}
Consider now the extensions described in \Fig{cap:fields-extension}.
From \Eq{eq:x_existence}, by considering the left extension of
\Fig{cap:fields-extension}, we deduce that \begin{equation}
  N_{\mathbb{K}/\mathbb{Q}(i)}(x)=N_{\mathbb{Q}\left(\zeta_{8}\right)/\mathbb{Q}(i)}\left(N_{\mathbb{K}/\mathbb{Q}\left(\zeta_{8}\right)}(x)\right)=\zeta_{8}\cdot\tau\left(\zeta_{8}\right)=-i.\label{eq:left-extension}\end{equation}
Now, we deduce, from the right extension of figure
\ref{cap:fields-extension} that \begin{equation}
  N_{\mathbb{K}/\mathbb{Q}(i)}(x)=N_{\mathbb{Q}\left(i,\sqrt{5}\right)/\mathbb{Q}(i)}\left(N_{\mathbb{K}/\mathbb{Q}\left(i,\sqrt{5}\right)}(x)\right)=-i.\label{eq:right-extension}\end{equation}
Denote
$y=N_{\mathbb{K}/\mathbb{Q}\left(i,\sqrt{5}\right)}(x)\in\mathbb{Q}\left(i,\sqrt{5}\right)$.
Then, the number $z=\frac{1+\sqrt{5}}{2}\cdot y$ has an algebraic norm
equal to $i$, and belongs to $\mathbb{Q}\left(i,\sqrt{5}\right)$.  In
\cite{Belfiore_golden}, it has been proved that $i$ was not a norm in
$\mathbb{Q}\left(i,\sqrt{5}\right)$.  So, $\zeta_{8}$ is not a norm in
$\mathbb{K}$.

\section{$\zeta_{16}$ is not a norm in $\mathbb{Q}\left(\zeta_{16},\sqrt{5}\right)$\label{zeta16-not-norm}}
The proof is similar to the one of appendix \ref{zeta8-not-norm}.
First, we assume that $\zeta_{16}$ is a norm in
$\mathbb{K}=\mathbb{Q}\left(\zeta_{16},\sqrt{5}\right)$, \ie,
\begin{equation} \exists
  x\in\mathbb{K},N_{\mathbb{K}/\mathbb{Q}\left(\zeta_{16}\right)}(x)=\zeta_{16}.\label{eq:x-existence-16}\end{equation}
We deduce that \begin{equation}
  N_{\mathbb{K}/\mathbb{Q}(i)}(x)=N_{\mathbb{Q}\left(\zeta_{16}\right)/\mathbb{Q}(i)}\left(N_{\mathbb{K}/\mathbb{Q}\left(\zeta_{16}\right)}(x)\right)=\zeta_{16}\cdot\tau\left(\zeta_{16}\right)\cdot\tau^2\left(\zeta_{16}\right)\cdot\tau^3\left(\zeta_{16}\right)=-i.\label{eq:left-extension-16}\end{equation}
But we also have, \begin{equation}
  N_{\mathbb{K}/\mathbb{Q}(i)}(x)=N_{\mathbb{Q}\left(i,\sqrt{5}\right)/\mathbb{Q}(i)}\left(N_{\mathbb{K}/\mathbb{Q}\left(i,\sqrt{5}\right)}(x)\right)=-i.\label{eq:right-extension-16}\end{equation}
Denote
$y=N_{\mathbb{K}/\mathbb{Q}\left(i,\sqrt{5}\right)}(x)\in\mathbb{Q}\left(i,\sqrt{5}\right)$.
Then the number $z=\frac{1+\sqrt{5}}{2}\cdot y$ has an algebraic
norm equal to $i$ and belongs to $\mathbb{Q}\left(i,\sqrt{5}\right)$, which is a contradiction.

\section*{Acknowledgement}
\label{sec:acknowledgement}
The authors would like to thank the anonymous reviewers for their
valuable comments.

\newpage

\newpage

\newpage

\newpage

\newpage

\end{document}